\documentclass[a4paper, 12pt]{article}

\pdfoutput = 1

\usepackage{styleBuding}
\usepackage{bm}
\usepackage{color,listings}
\usepackage[usenames,dvipsnames,svgnames,table]{xcolor}
\usepackage{amsthm, amsmath}
\usepackage{amssymb}
\usepackage{mathrsfs}
\usepackage{graphicx}
\usepackage{slashed}
\usepackage{soul}
\usepackage{multirow}
\usepackage{enumitem}
\usepackage{afterpage}
\usepackage{subfigure}
\usepackage{float}
\usepackage{diagbox}
\usepackage{siunitx} 
\usepackage{url} 
\usepackage[utf8]{inputenc}
\usepackage[figuresright]{rotating}
\definecolor{back}{HTML}{F8F8F8}
\usepackage{epstopdf}
\usepackage{lipsum}

\preprint{$\begin{gathered}\includegraphics[width=0.05\textwidth]{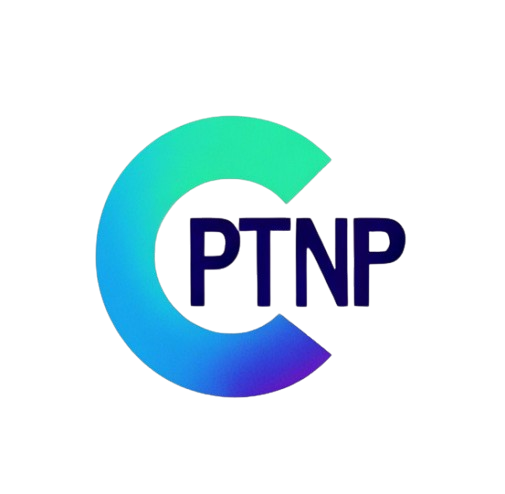}\end{gathered}$\, CPTNP-2025-040}

\newcommand\blfootnote[1]{%
	\begingroup
	\renewcommand\thefootnote{}\footnote{#1}%
	\addtocounter{footnote}{-1}%
	\endgroup
}

\newcommand{\rom}[1]{\uppercase\expandafter{\romannumeral #1\relax}}

\allowdisplaybreaks
\DeclareUnicodeCharacter{2212}{-}

\title{\boldmath Natural Realization of Tens-of-GeV Dark Matter in the GNMSSM}

\author{Fei Li$^{a}$, Junjie Cao$^{a,b,*}$\blfootnote{*Corresponding author.} }
\affiliation{$^a$ School of Physics, Henan Normal University, Xinxiang 453007, China}
\affiliation{$^b$ School of Physics, Zhengzhou University, Zhengzhou 450000, China}

\emailAdd{hnufeili@163.com}
\emailAdd{junjiec@alumni.itp.ac.cn}

\abstract{
This study presents a comparative analysis of the Minimal Supersymmetric Standard Model (MSSM), the $Z_3$-symmetric Next-to-Minimal Supersymmetric Standard Model ($Z_3$-NMSSM), and the General Next-to-Minimal Supersymmetric Standard Model (GNMSSM), incorporating constraints from dark matter (DM) relic density, the LUX-ZEPLIN 2024 experiment (LZ 2024), Higgs data, and the Large Hadron Collider (LHC). The results suggest that, among the three frameworks, only GNMSSM can naturally accommodate for light DM with a mass below $100~{\rm GeV}$. As such, the viable supersymmetry candidate is primarily Singlino-like. One key advantage of the GNMSSM is the effective decoupling between interactions that establish the relic density and those that control direct detection, allowing the model to satisfy all current experimental bounds simultaneously.
We further explore two characteristic mass hierarchies in the GNMSSM parameter space, each exhibiting distinct phenomenological behaviors.
The first hierarchy, $\tilde{S} < \tilde{B} < \tilde{H}$ (Singlino–Bino–Higgsino), involves a relatively light Bino and allows the Higgsino mass parameter, $\mu_{\rm tot}$, to be as low as about $200~{\rm GeV}$, naturally yielding light DM at tens of GeV. The dominant annihilation channels are then $\tilde{\chi}_1^0\tilde{\chi}_1^0 \to A_sA_s$ in the $h_1$ scenario and $\tilde{\chi}_1^0\tilde{\chi}_1^0 \to h_sA_s$ in the $h_2$ scenario, where $h_s$ and $A_s$ denote singlet-dominated CP-even and CP-odd Higgs bosons, respectively. The second hierarchy, $\tilde{S} < \tilde{H} < \tilde{B}$, corresponds to a heavy Bino. In this case, although the DM phenomenology remains qualitatively similar, LHC constraints require $\mu_{\rm tot} \gtrsim 900~{\rm GeV}$, implying a significant degree of fine-tuning in reproducing the $Z$-boson mass.
}

\begin{document}
\maketitle
\flushbottom

\section{Introduction}

Among numerous dark matter (DM) candidates, Weakly Interacting Massive Particles (WIMPs) have long been favored by theorists due to their natural ability to produce the observed cosmic abundance through the thermal freeze-out mechanism~\cite{Jungman:1995df}. According to this mechanism, WIMPs should interact with ordinary matter at a well-defined strength, yielding characteristic cross-sections: $\sim$$10^{-45}~{\rm cm^2}$ for spin-independent (SI) nucleon scattering~\cite{Baum:2017enm} and $\sim$$10^{-40}~{\rm cm^2}$ for spin-dependent (SD) interactions~\cite{Cao:2019qng}. However, direct detection experiments, particularly LUX-ZEPLIN 2024 (LZ 2024)~\cite{LZ:2024zvo}, have pushed these upper limits down to  $10^{-48}~{\rm cm^2}$ and  $10^{-43}~{\rm cm^2}$, respectively, posing substantial challenges for conventional WIMP models. To naturally resolve this tension, recent research has increasingly focused on extensions of economical WIMP theories by constructing DM structures that are nearly decoupled from the Standard Model (SM), leading to so-called ``Secluded dark matter'' models~\cite{Pospelov:2007mp}.

Cosmological measurements of effective light neutrino numbers and theoretical unitarity constraints place WIMPs within a broad mass range of $3.5~{\rm MeV}$~\cite{Ho:2012ug} to $100~{\rm TeV}$~\cite{Griest:1989wd,Smirnov:2019ngs}.  Within this window, the tens-of-GeV regime deserves special scrutiny for four reasons:
\begin{itemize}
\item Many DM theories link DM to electroweak symmetry breaking, providing theoretical motivation for DM at the electroweak scale~\cite{Feng:2010gw,Arcadi:2017kky}.
\item Direct detection experiments reach their highest sensitivity in this mass range~\cite{LZ:2024zvo}. Thus, understanding how a WIMP can evade current bounds is invaluable for model building.
\item Achieving the correct relic density often requires additional new particles with masses close to that of the WIMP. As such, establishing whether such states can elude Large Hadron Collider (LHC) searches is a key model-building constraint~\cite{Feng:2010gw,Arcadi:2017kky,Abdallah:2015ter}.
\item Most critically, after accounting for current experimental limits, many theories face severe fine-tuning problems when predicting DM in this mass range (see the discussion below). Should future experiments confirm DM in this region, a number of theoretical frameworks would face fundamental challenges. Therefore, systematically exploring models that naturally predict DM in this range, and distinguishing them from other theoretical frameworks, constitutes an important research direction at the intersection of particle physics and cosmology.
\end{itemize}

As one of the most compelling physical models beyond the SM, supersymmetry (SUSY) naturally accommodates WIMP DM~\cite{Jungman:1995df}. Consequently, studying tens-of-GeV DMs within supersymmetric frameworks has attracted considerable attention~\cite{Bottino:2002ry,Bottino:2003cz,Belli:2011kw,Gunion:2005rw}. This paper systematically compares the Minimal Supersymmetric Standard Model (MSSM)~\cite{Haber:1984rc,Gunion:1984yn} and the Next-to-Minimal Supersymmetric Standard Model with $Z_3$ symmetry ($Z_3$-NMSSM)~\cite{Ellwanger:2009dp,Miller:2003ay}, highlighting their characteristics and differences for predicting DM of several tens of GeV. We further demonstrate how the General NMSSM (GNMSSM) can predict a much richer spectrum of DM phenomena and discuss implications for future experiments.

The remainder of this paper is organized as follows. Section \ref{theory-section} highlights the similarities and differences between the MSSM, the $Z_3$-NMSSM, and the GNMSSM, in the context of attempts to realize a DM mass of ${\cal{O}} (10~{\rm GeV})$. In this process, the first two models are demonstrated to suffer from severe fine-tuning issues, while the GNMSSM avoids this problem. Section \ref{sec:NR} then focuses on the GNMSSM, examining in detail how it produces a viable tens-of-GeV DM compatible with all current constraints. Conclusions are then summarized in Section \ref{conclusion}.

\section{\label{theory-section} SUSY predictions of light DM} 
\subsection{MSSM }

In the MSSM, a compelling DM candidate in the tens-of-GeV mass range is the lightest neutralino, which is predominantly Bino-like. This particle couples to ordinary matter chiefly through its interactions with the Higgs sector and the $Z$ boson. The corresponding effective couplings are given by~\cite{Baum:2017enm}:\footnote{For mostly Bino-like DM with small Higgsino admixtures, these couplings give (to a good approximation) the same numerical results as Eqs.~(2.4)-(2.6) in Ref.~\cite{He:2023lgi}. The differences between the two may be understood as originating from corrections due to higher dimensional operators in Ref.~\cite{Baum:2017enm}, associated with the expansion of the denominator $\left[1-\left(M_1/\mu\right)^2 \right]$ in powers of $M_1^2/\mu^2$. }
\begin{eqnarray} \label{eq:gBH1}
	C_{\widetilde{B}\widetilde{B} H_{\rm SM}} &= & g_1^2 (\sin 2\beta + \frac{M_1}{\mu})\frac{v}{\mu}~,  \quad \quad C_{\widetilde{B}\widetilde{B} H_{\rm NSM}} =  g_1^2 \cos 2\beta \frac{v}{\mu}~, \\
	-i C_{\widetilde{B}\widetilde{B} A_{\rm NSM}} &= & g_1^2 ( 1 + \sin 2\beta \frac{M_1}{\mu})\frac{v}{\mu} ~, \quad \quad C_{\widetilde{B}\widetilde{B} Z} =  \frac{g_1^3 \cos 2\beta}{2 \sin \theta_W} \frac{v^2}{\mu^2}, \label{eq:gBH-1}
\end{eqnarray}
where $v=174~{\rm GeV}$ is the electroweak vacuum expectation value (vev), $|M_1|$ represents the DM mass, $H_{\rm SM} \equiv \sin\beta {\rm Re}(H_u^0) + \cos\beta {\rm Re} (H_d^0)$  corresponds to the SM Higgs field, and $H_{\rm NSM} \equiv \cos\beta {\rm Re}(H_u^0) - \sin\beta {\rm Re}(H_d^0)$ and $A_{\rm NSM} \equiv \cos\beta {\rm Im}(H_u^0) + \sin\beta  {\rm Im}(H_d^0)$ represent the additional doublet neutral Higgs fields.
These interactions give rise to both SI and SD scattering off nucleons. The corresponding cross-sections can be approximated by~\cite{Huang:2014xua,Cao:2019qng,He:2023lgi}:
\begin{eqnarray}
\sigma_{\tilde{\chi}_1^0,N}^{\rm SI} & & \simeq  2 \times 10^{-48} {\rm cm^2} \times \left ( \frac{F_u^N + F_d^N}{0.28} \right )^2 \times \left \{ \frac{F_u^N}{F_u^N + F_d^N} \right . \nonumber \\
& & \times  \left [  \frac{\cos \alpha}{\sin \beta} \left ( \frac{C_{\tilde{\chi}_1^0 \tilde{\chi}_1^0 h}}{0.002} \right ) \left ( \frac{125~{\rm GeV}}{m_h} \right )^2 +  \frac{\sin \alpha}{\sin \beta} \left ( \frac{C_{\tilde{\chi}_1^0 \tilde{\chi}_1^0 H}}{0.002} \right ) \left ( \frac{125~{\rm GeV}}{m_H} \right )^2 \right ] +  \frac{F_d^N}{F_u^N + F_d^N}   \nonumber \\
& & \left . \times \left [  - \frac{\sin \alpha}{\cos \beta} \left ( \frac{C_{\tilde{\chi}_1^0 \tilde{\chi}_1^0 h}}{0.002} \right ) \left ( \frac{125~{\rm GeV}}{m_h} \right )^2 +  \frac{\cos \alpha}{\cos \beta} \left ( \frac{C_{\tilde{\chi}_1^0 \tilde{\chi}_1^0 H}}{0.002} \right ) \left ( \frac{125~{\rm GeV}}{m_H} \right )^2 \right ] \right \}^2, \nonumber \\
& \simeq &  2 \times 10^{-48} {\rm cm^2} \times \nonumber \\
& & \left [ \left ( \frac{C_{\widetilde{B}\widetilde{B} H_{\rm SM}}}{0.002} \right )  \left ( \frac{125~{\rm GeV}}{m_h} \right )^2 - \cot 2 \beta \left (\frac{C_{\widetilde{B}\widetilde{B} H_{\rm NSM}}}{0.002} \right ) \left ( \frac{125~{\rm GeV}}{m_H} \right )^2 \right ]^2, \label{SI-1}  \\
\sigma_{\tilde{\chi}_1^0,N}^{\rm SD} && \simeq  C_N \times \left ( \frac{C_{\tilde{\chi}_1^0 \tilde{\chi}_1^0 Z}}{0.1} \right )^2. \label{SD-1}
\end{eqnarray}
In these expressions, $h$ corresponds to the scalar discovered at the LHC, $H$ denotes heavy doublet-dominated Higgs boson, and $F_u^N$ and $F_d^N$ represent the normalized contributions of up- and down-type quarks to the nucleon mass, respectively. The terms  $C_{\tilde{\chi}_1^0 \tilde{\chi}_1^0 h}$, $C_{\tilde{\chi}_1^0 \tilde{\chi}_1^0 H}$, and $C_{\tilde{\chi}_1^0 \tilde{\chi}_1^0 Z}$ quantify the DM couplings to the Higgs bosons $h$ and $H$, and the $Z$-boson, respectively. The coefficient $C_N$ relates to the nucleon spin, with values of $C_p \simeq 1.8 \times 10^{-40}~{\rm cm^2} $ for protons and
$C_n \simeq 1.4 \times 10^{-40}~{\rm cm^2} $ for neutrons. The angle $\alpha$ represents the mixing angle between the CP-even Higgs fields $H_{\rm SM}$ and $H_{\rm NSM}$ in forming the mass eigenstates $h$ and $H$, which satisfies the relationship $\alpha \simeq \beta - \pi/2$ in the large $m_A$ limit~\cite{Djouadi:2005gj}. This approximation was employed in deriving the final expression for $\sigma_{\tilde{\chi}_1^0,N}^{\rm SI}$.

The formulations presented above reveal important suppression mechanisms for DM-nucleon interactions. Specifically, as the Higgsino mass parameter $\mu$ increases, the Bino-Higgs coupling becomes suppressed by a factor of $1/\mu$, while the Bino-$Z$ coupling experiences an even stronger suppression by a factor of $1/\mu^2$. As a direct consequence, the SI scattering cross-section $\sigma_{\tilde{\chi}_1^0,N}^{\rm SI}$ is suppressed by $1/\mu^2$, while the SD scattering cross-section $\sigma_{\tilde{\chi}_1^0,N}^{\rm SD}$ undergoes a more dramatic reduction by a factor of $1/\mu^4$. In addition, a remarkable cancellation effect occurs when $\sin 2\beta = -M_1/\mu$, causing  $C_{\widetilde{B}\widetilde{B} H_{\rm SM}} = 0$ to vanish completely. This cancellation leads to a substantial suppression of the SI cross-section  $\sigma_{\tilde{\chi}_1^0,N}^{\rm SI}$~\cite{Huang:2014xua}.

Before analyzing DM properties in detail, key experimental constraints on the MSSM parameter space can be summarized as follows:
\begin{itemize}
\item To naturally accommodate the SM-like Higgs boson mass of approximately $125~{\rm GeV}$, the theory requires $\tan \beta \gtrsim 4$~\cite{Bagnaschi:2017tru}.
\item LHC searches for additional Higgs bosons indicate that if the heavy CP-even Higgs boson ($H$) decays exclusively into SM particles, its mass must exceed $350~{\rm GeV}$, regardless of $\tan \beta$ values~\cite{CMS:2022goy}.
\item  Compressed spectrum scenarios, where DM coannihilates with electroweakinos to achieve the observed relic abundance, are constrained by LHC experiments. These constraints establish lower mass bounds of approximately 180 GeV for Higgsinos and 220 GeV for Winos~\cite{ATLAS:2021moa}.
\item LHC searches for electroweakinos via multi-lepton and fully hadronic final states set stringent limits within simplified model frameworks: Higgsino mass limits reach $\sim 900~{\rm GeV}$ for massless DM, while Wino limits extend to $1200~{\rm GeV}$~\cite{ATLAS:2021moa,ATLAS:2021yqv}. These limits increase further when light sleptons are present in the decay chain~\cite{ATLAS:2024fub}.
\end{itemize}

Light DM scenarios in the MSSM exhibit the following salient features after integrating theoretical formulations with the experimental results:
\begin{itemize}
\item The branching ratio for $H$ decaying into DM pairs is significantly suppressed for large $\mu$ values. Neglecting this small channel, $m_H$ should exceed 350 GeV. In this regime, the correct relic density can only be obtained through $Z$- or $h$-funnel annihilation, requiring $|M_1| \simeq m_Z/2$ or $|M_1| \simeq m_h/2$~\cite{He:2023lgi}.
\item LZ 2024 constraints on $\sigma_{\tilde{\chi}_1^0,N}^{\rm SD}$ indicate that for $\tan \beta \gtrsim 4$, the $\mu$ parameter must exceed $700~{\rm GeV}$, as demonstrated by Eq.~(\ref{SD-1}). Furthermore, analyzing SI cross-section constraints with corresponding formula yields more nuanced boundaries that are dependent on the sign of $M_1/\mu$. Specifically, when $m_A=1~{\rm TeV}$, $\mu$ must exceed $900~{\rm GeV}$ in scenarios where $M_1/\mu$ is negative and $\tan \beta \lesssim 9$, or where $M_1/\mu$ is positive and $\tan \beta \lesssim 25$. These $\tan \beta$ thresholds increase modestly to 10 and 27, respectively, when $m_A = 2~{\rm TeV}$. Notably, these direct-detection bounds on $\mu$ are independent of Higgsino decay modes and therefore complement the LHC limits.
\item The Monte-Carlo simulations presented in this work suggest that for DM annihilation via $Z$ or Higgs resonance, LHC electroweakino searches require $\mu \gtrsim 900~{\rm GeV}$ when Higgsino decay into $H$ is kinematically forbidden.
\end{itemize}

Based on the preceding analysis, the MSSM encounters significant fine-tuning challenges when accommodating the light DM scenario:
\begin{itemize}
\item According to the $Z$-boson mass expression~\cite{Baer:2012uy}:
\begin{eqnarray}
m^2_{Z}=\frac{2 (m^2_{H_d}+\Sigma_{d})- 2 (m^2_{H_u}+
\Sigma_{u})\tan^{2}\beta}{\tan^{2}\beta-1}- 2 \mu^{2},
\label{minimization}
\end{eqnarray}
approximately $m_Z^2/(2 \mu^2) \simeq 0.8\%$ parameter fine-tuning is required to predict the correct $Z$ boson mass  given that $\mu \gtrsim 700 ~{\rm GeV}$~\cite{Baer:2013gva}.
\item Since large $\mu$ values severely suppress DM couplings to $Z$ and Higgs bosons, achieving the correct cosmic abundance requires the DM mass to approach resonance conditions ($m_Z/2$ or $m_h/2$) with extreme precision, introducing another significant fine-tuning issue~\cite{Cao:2018rix}.
\end{itemize}
It is worth emphasizing that both instances of fine-tuning arise from a single theoretical demand: the 
$\mu$-parameter must be comparatively large. This large value, in turn, heavily suppresses all DM couplings.

\subsection{$Z_3$-NMSSM}\label{sec:z3}

In the $Z_3$-NMSSM, DM with a mass of several tens of GeV can either be a Bino-dominated lightest neutralino or a Singlino-dominated lightest neutralino. In addition to the couplings shown in Eq.~(\ref{eq:gBH-1}), the interactions listed below also influence DM phenomenology~\cite{Baum:2017enm}:
\begin{eqnarray}
C_{\widetilde{B}\widetilde{B} {\rm Re}[S]} &=& i C_{\widetilde{B}\widetilde{B} {\rm Im}[S]} = - \frac{\lambda g_1^2 \sin 2\beta}{2} \frac{v^2}{\mu_{\rm eff}^2}, \nonumber \\
C_{\widetilde{B}\widetilde{S} H_{\rm SM}} & = & - 2 \sqrt{2} \lambda g_1 \cos 2\beta \frac{v}{\mu_{\rm eff}}, \quad 	
i C_{\widetilde{B}\widetilde{S} A_{\rm NSM}} = - \sqrt{2} \lambda g_1 \cos 2\beta \frac{\left( M_1 + M_{\widetilde{S}} \right) v }{\mu_{\rm eff}^2}, \nonumber \\
C_{\widetilde{B}\widetilde{S} H_{\rm NSM}} &=& \sqrt{2} \lambda g_1 \left [2 \sin 2\beta  + \frac{ \left(M_1 - M_{\widetilde{S}} \right)}{\mu_{\rm eff}} \right ]\frac{v}{\mu_{\rm eff}}, \nonumber   \\
C_{\widetilde{B}\widetilde{S} {\rm Re}[S]} & = & i C_{\widetilde{B}\widetilde{S} {\rm Im}[S]} = \sqrt{2} \lambda^2 g_1 \cos 2\beta \frac{v^2}{\mu_{\rm eff}^2}, \quad  C_{\widetilde{B}\widetilde{S} Z} =  \frac{\sqrt{2} \lambda g_1^2 \sin 2 \beta}{\sin \theta_W} \frac{v^2}{\mu_{\rm eff}^2}, \nonumber  \\
 C_{\widetilde{S}\widetilde{S} H_{\rm SM}} &=& -  2 \lambda^2 \sin 2\beta \frac{v}{\mu_{\rm eff}} + 2 \lambda^2 
\frac{m_{\tilde{S}} v}{\mu_{\rm eff}^2}, \quad \quad C_{\widetilde{S}\widetilde{S} H_{\rm NSM}} = - 2 \lambda^2 \cos 2\beta \frac{v}{\mu_{\rm eff}},  \nonumber \\
C_{\widetilde{S}\widetilde{S} {\rm Re}[S]} &=& i C_{\widetilde{S}\widetilde{S} {\rm Im}[S]} = \lambda^3 \sin 2\beta \frac{v^2}{\mu_{\rm eff}^2} - 2 \kappa \left( 1 - \lambda^2 \frac{v^2}{\mu_{\rm eff}^2} \right), \nonumber \\ 
 i C_{\widetilde{S}\widetilde{S} A_{\rm NSM}} &= & 2 \lambda^2  \left [ 1 - \sin 2 \beta \frac{m_{\tilde{S}}}{\mu_{\rm eff}} \right ]\frac{v}{\mu_{\rm eff}}, \quad \quad C_{\widetilde{S}\widetilde{S} Z} = - \frac{\lambda^2 g_1  \cos 2 \beta}{\sin \theta_W}  \frac{v^2}{\mu_{\rm eff}^2},  \label{NMSSM-Couplings}
\end{eqnarray}
where $\mu_{\rm eff} \equiv \lambda \langle S\rangle $ is the effective Higgsino mass, and $M_{\tilde{S}} \equiv 2 \kappa  \mu_{\rm eff}/\lambda$ is the Singlino mass. Unlike the interactions in Eq.~(\ref{eq:gBH-1}), these couplings depend not only on the parameter $\mu_{\rm eff}$ but also critically on the Higgs field Yukawa coupling coefficients ($\lambda$ and $\kappa$). When analyzing their impact on DM phenomenology, three fundamental constraints must be considered, as described below:
\begin{itemize}
\item Higgs Data Fitting

In the NMSSM, the singlet CP-even Higgs field mixes with the SM Higgs field to form mass eigenstates, consequently modifying the properties of the SM-like Higgs particle. This effect becomes particularly evident when diagonalizing the CP-even Higgs mass matrix. Specially, in the basis ($H_{\rm NSM}$, $H_{\rm SM}$, ${\rm Re}[S]$), the mass matrix takes the following form~\cite{Ellwanger:2009dp,Miller:2003ay}:
 \begin{eqnarray}
  {\cal M}_{S, 11}^2 &=&  M^2_A + (m^2_Z -\lambda^2 v^2) \sin^2 2\beta, \quad {\cal M}_{S,12}^2 =  -\frac{\sin 4 \beta}{2}(m^2_Z-\lambda^2 v^2), \nonumber \\
  {\cal M}_{S,13}^2 &=&  - \lambda \cos2\beta (\frac{M^2_A \sin 2 \beta}{2\mu_{\rm eff}}+\frac{\kappa \mu_{\rm eff}}{\lambda}) v, \quad   {\cal M}_{S, 22}^2 =  m_Z^2\cos^2 2\beta +\lambda^2v^2\sin^2 2\beta, \nonumber \\
  {\cal M}_{S, 23}^2 &=&  2\lambda [1-(\frac{M_A \sin 2 \beta}{2\mu_{\rm eff}})^2 -\frac{\kappa}{2\lambda}\sin2\beta] \mu_{\rm eff} v \equiv 2 \lambda \delta \mu_{\rm eff} v, \nonumber \\
  {\cal M}_{S, 33}^2 &=& \lambda^2 (\frac{M_A \sin 2 \beta }{2 \mu_{\rm eff}})^2 v^2 + \frac{\kappa}{\lambda} \mu_{\rm eff} A_{\kappa} + 4 \left (\frac{\kappa}{\lambda} \right )^2 \mu_{\rm eff}^2 -\frac{1}{2} \lambda\kappa \sin 2\beta v^2\equiv m^2_B, \nonumber
\end{eqnarray}
where the parameter $\delta$ functions as a measure of the cancellation between different terms in ${\cal M}_{S, 23}^2$~\cite{Cao:2012fz}. 
The three CP-even mass eigenstates $h_i=\{h, H, h_s\}$  are then acquired through unitary rotations of the matrix $V$ that diagonalizes ${\cal{M}}_S^2$, yielding:
\begin{eqnarray} \label{Mass-eigenstates}
    h_i & = & V_{h_i}^{\rm NSM} H_{\rm NSM}+V_{h_i}^{\rm SM} H_{\rm SM}+V_{h_i}^{\rm S} {\rm Re}[S],  \\
    C_{\tilde{\chi} \tilde{\chi} h_i} & = &  V_{h_i}^{\rm NSM} C_{\tilde{\chi} \tilde{\chi} H_{\rm NSM}} + V_{h_i}^{\rm SM} C_{\tilde{\chi} \tilde{\chi} H_{\rm SM}} +V_{h_i}^{\rm S} C_{\tilde{\chi} \tilde{\chi} {\rm Re}[S]}, \quad \tilde{\chi} = \tilde{B}, \tilde{S}.  
\end{eqnarray}
Two scenarios can then be further distinguished based on the mass ordering of these CP-even states: the $h_1$ scenario with $m_h < m_{h_s} < m_H$, and the $h_2$ scenario with $m_{h_s} < m_h < m_H$. 
The following approximations are valid in cases with a very massive $H$~\cite{Baum:2017enm}:
\begin{eqnarray}
\frac{V_{h}^{\rm S}}{V_h^{\rm SM}} & \simeq &  \frac{{\cal M}^2_{S, 23}}{(m_h^2 -  {\cal M}_{S, 33}^2)}, \quad V_{h}^{\rm NSM} \sim 0, \quad V_h^{\rm SM} \simeq \sqrt{1 - \left ( \frac{V_{h}^{\rm S}}{V_h^{\rm SM}} \right )^2}  \sim 1. \label{Approximations}
\end{eqnarray}
The latest HiggsTools package~\cite{Bahl:2022igd}, used to conduct a comprehensive global fit of Higgs properties, then determined that LHC Higgs data constrains
$ |{\cal M}^2_{S, 23}/(m_h^2 -  {\cal M}_{S, 33}^2)| \lesssim 0.4$~\cite{Cao:2024axg}, or equivalently $\lambda \delta \mu_{\rm tot} \lesssim | {\cal M}_{S, 33}^2 - m_h^2|/(900~{\rm GeV})$. 
Quantitatively, $\lambda \lesssim 0.009$ $ (0.039)$ when ${\cal{M}}_{S, 33} = 100 ~{\rm GeV}$ (200~{\rm GeV}), $\mu_{\rm eff} = 700~{\rm GeV}$, and $\delta \simeq 1$. Furthermore, maintaining perturbativity of the theory up to the GUT scale imposes the additional requirement that $\lambda \lesssim 0.7$~\cite{Ellwanger:2009dp}.
 
\item Singlino Mass Constraint

Assuming that the Singlino plays a significant role in DM physics, its mass given by  $m_{\tilde{S}} = 2 \kappa/\lambda \mu_{\rm eff}$ should be around several tens of GeV. This relationship implies:
\begin{eqnarray}
\kappa \lesssim \frac{\lambda}{20} \times \frac{m_{\tilde{S}}}{100~{\rm GeV}} \times \frac{1~{\rm TeV}}{\mu_{\rm eff}}. \label{Bound-kappa}
\end{eqnarray}
Consequently, in light DM scenarios, the parameter $\kappa$ must be at least one order of magnitude smaller than $\lambda$.

\item DM-nucleon Scattering

In the NMSSM framework, the SI cross-section for DM-nucleon scattering can be approximated by~\cite{Baum:2017enm}:
\begin{eqnarray}
\sigma_{\tilde{\chi},N}^{\rm SI} & \simeq &  2 \times 10^{-48} {\rm cm^2} \times \left [ \left ( \frac{125~{\rm GeV}}{m_h} \right )^2 \left (\frac{C_{\tilde{\chi} \tilde{\chi} h}}{0.002} \right ) \right . \nonumber \\
& & \left . +  \left ( \frac{125~{\rm GeV}}{m_H} \right )^2 \left (\frac{C_{\tilde{\chi} \tilde{\chi} H}}{0.002} \right ) + \left ( \frac{125~{\rm GeV}}{m_{h_s}} \right )^2 \left (\frac{C_{\tilde{\chi} \tilde{\chi} h_s}}{0.002} \right )  \right ]^2,
\end{eqnarray}
where the three terms correspond to exchanges of the SM-like Higgs $h$, the heavier doublet-like Higgs $H$, and the singlet-like Higgs $h_s$, respectively. Meanwhile, the SD cross-section maintains the same form as previously expressed in Eq.~(\ref{eq:gBH-1}). In the specific case of a Singlino-like DM, more detailed expressions have been thoroughly derived for both cross-sections in Refs.~\cite{Zhou:2021pit,Cao:2021ljw,Meng:2024lmi}. These derivations reveal important relationships, with the primary contribution to the SI cross-section scaling as $\sigma_{\tilde{\chi},N}^{\rm SI} \propto \lambda^4 v^2/\mu_{\rm eff}^2$ for a sufficiently massive $h_s$ or $\propto \lambda^2\kappa^2$ for a moderately light $h_s$, while the SD cross-section follows $\sigma_{\tilde{\chi},N}^{\rm SD} \propto \lambda^4 v^4/\mu_{\rm eff}^4$, highlighting the strong dependence of both cross-sections on the $\lambda$ parameter. LZ 2024 placed significant limitations on this parameter, restricting $\lambda$ to values as low as approximately $0.01$~\cite{Meng:2024lmi}. This constraint has profound implications for the viability of Singlino-like DM candidates within the NMSSM framework.
\end{itemize}

When DM is predominantly Bino-like, LZ 2024-based constraints on the SD cross-section require $\mu_{\rm eff} \gtrsim 700~{\rm GeV}$ for $\tan \beta \gtrsim 4$, as in the MSSM. In addition to the canonical $Z$ and $h$ resonant annihilation channels, DM may also co-annihilate with Singlino when the Bino and Singlino masses are nearly degenerate~\cite{Baum:2017enm}, or proceed through $h_s$ and $A_s$ resonant channels to achieve the experimentally measured relic abundance~\cite{Cao:2011re,Cao:2013gba,Cao:2018rix}. The relevant interactions — including  $C_{\tilde{B} \tilde{B} {\rm Re}[S]}$, $C_{\tilde{B} \tilde{B} {\rm Im}[S]}$, $C_{\tilde{B} \tilde{S} H_{\rm SM}}$,  $C_{\tilde{B} \tilde{S} {\rm Re}[S]}$, $C_{\tilde{B} \tilde{S} {\rm Im}[S]}$, $C_{\tilde{B} \tilde{S} Z}$, $C_{\tilde{S} \tilde{S} H_{\rm SM}}$,  $C_{\tilde{S} \tilde{S} {\rm Re}[S]}$, $C_{\tilde{S} \tilde{S} {\rm Im}[S]}$, and $C_{\tilde{S} \tilde{S} Z}$ — are all suppressed by a factor of $1/\mu_{\rm eff}$ and typically also by $\lambda$. This suppression necessitates fine-tuning to reproduce the observed relic density. 

For Singlino-dominated DM, LZ 2024 constraints on both SI and SD cross-sections require $\lambda \lesssim 0.05$, although definitive lower bounds cannot be established on $\mu_{\rm eff}$~\cite{Meng:2024lmi}. Complementarily, LHC constraints on compressed spectrum scenarios mandate $\mu_{\rm eff} \gtrsim 180~{\rm GeV}$~\cite{ATLAS:2021moa}, while LHC searches for electroweakinos via multi-lepton channels impose even stronger constraints in other scenarios~\cite{ATLAS:2021moa,ATLAS:2021yqv,ATLAS:2024fub}. As a result, all interactions in Eq.~(\ref{NMSSM-Couplings}) are significantly suppressed by  $\lambda$ or $\kappa$, where $\kappa \lesssim \lambda/4$ according to Eq.(\ref{Bound-kappa}) and the LHC restrictions on $\mu_{\rm eff}$. Under these circumstances, annihilation processes such as $\tilde{S} \tilde{S} \to h_s A_s, h_s h_s, A_s A_s$ cannot generate the observed relic density due to the extremely small value of $\kappa$. Conversely, resonant $h_s$ or $A_s$ annihilation requires substantial fine-tuning, since the coupling strengths $C_{\tilde{S} \tilde{S} {\rm Re}[S]}$ and $C_{\tilde{S} \tilde{S} {\rm Im}[S]}$ are weak~\cite{Meng:2024lmi}. 

The takeaway from the above discussion is clear: because every DM coupling is highly suppressed, the $Z_3$-NMSSM requires severe fine-tuning to accommodate a viable light DM candidate.
\subsection{GNMSSM }

In contrast to the $Z_{3}$-NMSSM, the GNMSSM drops the rigid $Z_{3}$ symmetry and incorporates all renormalizable operators in the singlet–Higgs sector. Beyond the familiar $\lambda\,\hat{S}\hat{H}_{u}\!\cdot\!\hat{H}_{d}$ and $\kappa\,\hat{S}^{3}/3$ terms, the superpotential includes an explicit Higgsino mass term $\mu\,\hat{H}_{u}\!\cdot\!\hat{H}_{d}$, a singlet Majorana mass term $\mu'\,\hat{S}^{2}/2$, and optionally a linear tadpole term $\xi\,\hat{S}$~\cite{Ellwanger:2009dp,Ross:2011xv}. These $Z_{3}$-violating terms simultaneously resolve the cosmological domain-wall problem~\cite{Abel:1996cr,Lee:2010gv,Lee:2011dya,Ross:2011xv,Ross:2012nr} and eliminate dangerous higher-order tadpoles~\cite{Ellwanger:1983mg, Ellwanger:2009dp}, while preserving theoretical motivation through discrete $R$-symmetries that naturally generate $\mu$ and $\mu'$ at the weak scale~\cite{Abel:1996cr,Lee:2010gv,Lee:2011dya,Ross:2011xv,Ross:2012nr}.
Although the Bino and Singlino interactions in the GNMSSM remain the same as those of the $Z_3$-NMSSM, the enlarged superpotential decouples several key mass parameters~\cite{Cao:2023gkc,Meng:2024lmi}. Specifically, the Higgsino mass $\mu_{\text{tot}}$ and the Singlino mass $m_{\tilde{S}}$ can be expressed as:
\begin{eqnarray}
\mu_{\text{tot}} = \lambda\langle S\rangle + \mu, \quad m_{\tilde{S}} = 2\kappa\langle S\rangle + \mu^\prime,
\end{eqnarray}
allowing them to be adjusted independently. These formulae represents a critical departure from the $Z_{3}$-NMSSM, in which the Singlino mass differs from the Higgsino mass only by a factor of $2\kappa/\lambda$, where $\kappa$ must be less than $\lambda/2$ for viable Singlino-like DM scenarios
and $\lambda/4$ if the DM is further required to be lighter than $100~{\rm GeV}$. In contrast, the GNMSSM permits $|\kappa|$ to be much larger than $\lambda$ in this scenario~\cite{Cao:2021ljw}. This freedom profoundly alters Singlino DM phenomenology while leaving Bino-like DM properties essentially unchanged. 

A Singlino-like DM $\tilde{\chi}^0_1$, the lightest supersymmetric particle (LSP) with a mass of $|m_{\tilde{\chi}^0_1}|\simeq |m_{\tilde{S}}|$, can thus annihilate through ``secluded sector'' channels controlled by $\kappa$, e.g., $\tilde{\chi}^0_1\tilde{\chi}^0_1 \rightarrow h_s A_s, h_s h_s, A_s A_s$. Consequently, the thermally averaged cross-section exhibits a characteristic $\kappa^4$ dependence~\cite{Meng:2024lmi}. In contrast, 
the direct detection cross-section, which is proportional to $\lambda^{4}$ or $\lambda^2\kappa^2$ like the $Z_3$-NMSSM, remains naturally suppressed due to the small value of $\lambda$~\cite{Meng:2024lmi}. This separation of roles offers a substantially broader DM parameter space, capable of simultaneously accommodating relic density observations and direct detection bounds, particularly obviating the need for the resonance enhancements in the annihilation~\cite{Cao:2021ljw,Meng:2024lmi}. In addition, smaller $\lambda$ values in viable regions enhance quartic self-coupling stability, thereby mitigating potential issues with color- or charge-breaking vacua that can plague certain regions of the $Z_3$-NMSSM parameter space~\cite{Hollik:2018yek,Hollik:2018wrr}.

From the collider perspective, the LHC phenomenology of GNMSSM may present both challenges and opportunities. For instance, the scalar and neutralino spectrum is enriched, as singlet scalar masses $m_{h_s}, m_{A_s}$ and the Singlino mass $m_{\tilde{S}}$ can be set independently, leading to diverse and potentially distinctive collider signatures~\cite{Cao:2023gkc}. Supersymmetric particle decay chains can be longer and more intricate, with significant momentum transfer to particles in the secluded sector~\cite{Cao:2021ljw}. While this may reduce the sensitivity of conventional searches, with Higgsino masses as low as about $200~{\rm GeV}$, it also opens new avenues for novel detection strategies~\cite{Cao:2022ovk}. 

In summary, the GNMSSM represents a well-motivated extension of the conventional $Z_3$-NMSSM, offering greater theoretical consistency and phenomenological flexibility. By introducing $\mu$, $\mu'$, and optionally $\xi$ terms that break the $Z_3$ symmetry, this enables $\lambda$ and $\kappa$ to play distinct roles in DM physics: $\lambda$ can remain minimal to satisfy direct detection constraints, while $\kappa$ can assume larger values to dominate DM self-annihilation processes. This characteristic makes the GNMSSM framework particularly attractive for addressing contemporary challenges in supersymmetry, naturalness, DM phenomenology, and Higgs physics in a coherent and less fine-tuned manner. As demonstrated below, this theory naturally predicts a viable DM candidate with a mass of several tens of GeV, since several key DM couplings remain appreciable even after all experimental constraints have been applied.

\section{Comprehensive study of GNMSSM}\label{sec:NR}

This section presents a comprehensive numerical framework for studying light DM in the GNMSSM. The model was built using the \textsf{SARAH-4.15.3}~\cite{Staub:2008uz, Staub:2012pb, Staub:2013tta, Staub:2015kfa} package. The  particle spectra and flavor observables were calculated by the \textsf{SPheno-4.0.5}~\cite{Porod:2003um,Porod:2011nf} and \textsf{FlavorKit}~\cite{Porod:2014xia} codes, respectively. DM observables were computed with \textsf{micrOMEGAs-5.0.4}~\cite{Belanger:2001fz, Belanger:2005kh, Belanger:2006is, Belanger:2010pz, Belanger:2013oya, Barducci:2016pcb} code. 

\subsection{Research strategy}\label{sec:RS}

A joint likelihood function $\mathcal{L}$ was first formulated to guide the sampling process and accurately delineate the GNMSSM parameter space as follows:
\begin{eqnarray}
\mathcal{L} & \equiv & \mathcal{L}_{\Omega h^2} \times \mathcal{L}_{\rm LZ} \times \mathcal{L}_{\rm Const},  \nonumber \\
\mathcal{L}_{\Omega h^2} &=& \exp\left[ -\frac{1}{2}\left( \frac{\Omega h^2-0.120}{0.012 } \right)^2 \right], \quad  \mathcal{L}_{\rm LZ}= \exp \left(-\frac{\sigma^{\rm SI}_{\rm eff}}{2\delta_{\sigma}^2} \right), \nonumber \\
\mathcal{L}_{\rm Const} &=& \left \{ \begin{aligned} & 1 & &{\rm if\ satisfying\ all\ experimental\ constraints} \\ &\exp\left(-100\right) & &{\rm otherwise} \end{aligned} \right .. \label{Likelihood}
\end{eqnarray}
In this construction, the relic abundance $\Omega h^2$  is assumed to follow a Gaussian distribution $\mathcal{L}_{\Omega h^2}$, with its central value derived from the Planck experiment~\cite{Planck:2018vyg} and a theoretical uncertainty of $10\%$. Given the smallness of the SI scattering cross-section and consequently $\sigma^{\rm SI}_p$ may significantly deviate from $\sigma^{\rm SI}_n$,
the effective SI cross-section in the Gaussian likelihood term ${\cal{L}}_{\rm LZ}$ is defined as $\sigma_{\rm eff}^{\rm SI} = 0.169\sigma_p^{\rm SI} + 0.347\sigma_n^{\rm SI} + 0.484\sqrt{\sigma_p^{\rm SI}\times\sigma_n^{\rm SI}}$~\cite{Cao:2019aam} for comarision with the results from LZ 2022~\cite{LZ:2022lsv}. The variance square is then given by $\delta_{\sigma}^2 \equiv {\rm UL}_\sigma^2/1.64^2 + (0.2\sigma)^2$, where ${\rm UL}_\sigma$ refers to the $90\%$ C. L. upper limit on the SI scattering cross-section, and $0.2 \sigma$ accounts for  uncertainties~\cite{Matsumoto:2016hbs}. The $\mathcal{L}_{\text{Const}}$ term encompasses the following experimental constraints:

\begin{itemize}
\item \textbf{Higgs data fit:} The properties of $h$ must align with the LHC Higgs data at $95\%$ C.L. This alignment is enforced by the \textsf{HiggsSignal 2.6.2} code~\cite{HS2013xfa,HSConstraining2013hwa,HS2014ewa,HS2020uwn} with the requirement that the $p$-value is greater than 0.05, where a 3 GeV combined theoretical and experimental uncertainty for $m_h$ is also considered. In addition, \textsf{HiggsTools-1.2}~\cite{Bahl:2022igd}, which incorporates 129 Higgs observables, was used to update constraints from the Higgs data by imposing the condition $\Delta\chi^2 \equiv \chi^2_{\text{Higgs}} - \chi^2_{\text{SM},125} \lesssim 6.18$, where $\chi^2_{\text{SM},125} \simeq 176.3$ corresponds to a pure SM Higgs boson in the GNMSSM~\cite{Cao:2024axg}.

\item \textbf{Extra Higgs searches:} The \textsf{HiggsBounds 5.10.2} code~\cite{HB2008jh,HB2011sb,HBHS2012lvg,HB2013wla,HB2020pkv} was employed to conduct comprehensive searches for additional Higgs bosons beyond the SM at LEP, Tevatron and LHC. This search was further complemented by \textsf{HiggsTools-1.2}~\cite{Bahl:2022igd} with enhanced search capabilities.

\item \textbf{Indirect DM searches:} The Fermi-LAT collaboration has accumulated years of dwarf galaxy observations, yielding constraints on DM annihilation cross-sections\footnote{see: www-glast.stanford.edu/pub\_data/1048}. The likelihood function from Refs.~\cite{Carpenter:2016thc,Huang:2016tfo} was applied to evaluate these bounds. The latest \textsf{MADHAT} package~\cite{Boddy:2019kuw, Boddy:2024tiu}, which integrates 14 years of public Fermi-LAT data from 54 dwarf spheroidal galaxies, was also employed to enhance the robustness of this analysis.
\item \textbf{$B$-physics observables:} The branching ratios of $B_s \to \mu^+ \mu^-$ and $B \to X_s \gamma$ should agree with experimental measurements at the $2\sigma$ level~\cite{pdg2018}.
\end{itemize}

\begin{table}[tbp]
\caption{An exploration of the parameter space for the $h_1$ scenario. All inputs were assumed to exhibit a flat prior distribution, reflecting their physical meaning in the small $\lambda$ case as shown in Ref.~\cite{Cao:2023gkc}. The parameter $v_s$ was defined as $\sqrt{2} \langle S\rangle$, and $m_B$ and $m_C$ are the mass parameters for the CP-even and -odd singlet Higgs fields, respectively. The soft trilinear coefficients for the third-generation squarks were set equal $(A_t = A_b)$ for simplicity and allowed to vary due to their significant radiative corrections to the SM-like Higgs boson mass. We fixed $M_1=150~{\rm GeV}$, $M_3=3~{\rm TeV}$, and set all other unmentioned dimensional parameters to $2~{\rm TeV}$. All parameters were defined at the renormalization scale $Q_{\rm input} = 1~{\rm TeV}$. The $h_2$ scenario shared the same parameter space as in the $h_1$ scenario, except that $ 5~{\rm GeV} \leq m_B \leq 130~{\rm GeV}$.
\label{tab:parameter-space}}
\centering
\vspace{0.3cm}
\resizebox{0.7\textwidth}{!}{
    \begin{tabular}{c|c|r@{\,--\,}l||c|c|r@{\,--\,}l}
      \hline
      \textbf{Parameter} & \textbf{Prior} & \multicolumn{2}{c||}{\textbf{Range}} & \textbf{Parameter} & \textbf{Prior} & \multicolumn{2}{c}{\textbf{Range}} \\
      \hline
      $\lambda$ & Flat & $10^{-3}$&0.2 & $\mu_{\rm tot}$/GeV & Flat & 200&1000\\
      $\kappa$ & Flat & -0.7&0.7 & $m_B$/GeV & Flat & 95&250 \\
      $\delta$ & Flat & -1.0&1.0 & $m_C$/GeV & Flat & 5&250 \\
      $\tan\beta$ & Flat & 1&60 & $m_{\tilde{S}}$/GeV & Flat & -102&102 \\ 
      $v_s$/GeV & Flat & 1&$10^3$ & ~$A_t$/TeV & Flat & -5.0&5.0 \\
      \hline
    \end{tabular}}
\end{table} 

\begin{table}[ht]
\caption{LHC analyses for electroweakino pair production included in this study, categorized by final-state signatures.}
\label{tab:LHC1}
\vspace{0.2cm}
\resizebox{0.96\textwidth}{!}{
\begin{tabular}{llll}
\hline\hline
\texttt{\textbf{Scenario}} & \texttt{\textbf{Final State}} &\multicolumn{1}{l}{\texttt{\textbf{Analysis}}}\\\hline		
\multirow{2}{*}{$\tilde{\chi}_{1}^{+}\tilde{\chi}_{1}^{-}\rightarrow 2\tilde{\ell}\nu(\ell\tilde\nu)$}&\multirow{2}{*}{$2\ell + \text{E}_\text{T}^{\text{miss}}$}&\texttt{ATLAS-CONF-2016-096($13.3fb^{-1}$)}~\cite{ATLAS:2016uwq}\\&&\texttt{ATLAS-1908-08215($139fb^{-1}$)}~\cite{ATLAS:2019lff}\\\\	
\multirow{1}{*}{$\tilde{\chi}_1^{+}\tilde{\chi}_1^{-}\rightarrow WW\tilde{\chi}_1^0 \tilde{\chi}_1^0$}&\multirow{1}{*}{$2\ell + \text{E}_\text{T}^{\text{miss}}$}&\texttt{ATLAS-1908-08215($139fb^{-1}$)}~\cite{ATLAS:2019lff}\\\\
\multirow{3}{*}{$\tilde{\chi}_{2}^{0}\tilde{\chi}_{1}^{\pm}\rightarrow \ell\tilde{\ell}\tilde{\ell}\nu(\ell\tilde\nu\nu\tilde\nu)$}&\multirow{3}{*}{$3\ell + \text{E}_\text{T}^{\text{miss}}$}&\texttt{ATLAS-CONF-2016-096($13.3fb^{-1}$)}~\cite{ATLAS:2016uwq}\\&& \texttt{ATLAS-1803-02762($36.1fb^{-1}$)}~\cite{ATLAS:2018ojr} \\&&\texttt{CMS-SUS-16-039($35.9fb^{-1}$)}~\cite{CMS:2017moi}\\\\
\multirow{2}{*}{$\tilde{\chi}_{2}^{0}\tilde{\chi}_{1}^{\pm}\rightarrow WZ\tilde{\chi}_1^0\tilde{\chi}_1^0$}&\multirow{2}{*}{$3\ell + \text{E}_\text{T}^{\text{miss}}$}&\texttt{ATLAS-2106-01676($139fb^{-1}$)}~\cite{ATLAS:2021moa}\\&&\texttt{ATLAS-CONF-2019-020($139fb^{-1}$)}~\cite{ATLAS:2019nnv}\\\\
$\tilde{\chi}_1^{0}\tilde{\chi}_2^{0}\rightarrow Z\tilde{\chi}_1^0\tilde{\chi}_1^0$&\multirow{4}{*}{$n\ell(n\geq4) + \text{E}_\text{T}^{\text{miss}}$}&\multirow{4}{*}{\texttt{ATLAS-2103-11684($139fb^{-1}$)}~\cite{ATLAS:2021yyr}}\\$\tilde{\chi}_1^{0}\tilde{\chi}_1^{\pm}\rightarrow W\tilde{\chi}_1^0\tilde{\chi}_1^0$&&\\$\tilde{\chi}_2^{0}\tilde{\chi}_1^{\pm}\rightarrow WZ\tilde{\chi}_1^0\tilde{\chi}_1^0$&&\\$\tilde{\chi}_1^{+}\tilde{\chi}_1^{-}\rightarrow WW\tilde{\chi}_1^0\tilde{\chi}_1^0$&&\\\\	
$\tilde{\chi}_2^0\tilde{\chi}_1^{\pm}\rightarrow\tau \tilde{\tau} \tilde{\tau}\nu$&$3\tau + \text{E}_\text{T}^{\text{miss}}$&\texttt{CMS-SUS-16-039($35.9fb^{-1}$)}~\cite{CMS:2017moi}\\\\ 	
 $\tilde{\chi}_2^0\tilde{\chi}_1^{\pm}\rightarrow \ell\tilde{\ell}\tilde{\tau}\nu$&$2\ell + 1\tau + \text{E}_\text{T}^{\text{miss}}$&\texttt{CMS-SUS-16-039($35.9fb^{-1}$)}~\cite{CMS:2017moi}\\\\
{$\tilde{\chi}_2^{0}\tilde{\chi}_1^{\pm}\rightarrow WZ\tilde{\chi}_1^0\tilde{\chi}_1^0$}&\multirow{2}{*}{$2j(\text{large}) + \text{E}_\text{T}^{\text{miss}}$}&\multirow{2}{*}{\texttt{ATLAS-2108-07586($139fb^{-1}$)}~\cite{ATLAS:2021yqv}}\\{$\tilde{\chi}_1^{+}\tilde{\chi}_1^{-}\rightarrow WW\tilde{\chi}_1^0\tilde{\chi}_1^0$}&&\\\\
{$\tilde{\chi}_2^{0}\tilde{\chi}_3^{0}\rightarrow ZZ(Zh)\tilde{\chi}_1^0\tilde{\chi}_1^0$}&\multirow{2}{*}{$j(\text{large}) + b(\text{large}) + \text{E}_\text{T}^{\text{miss}}$}&\multirow{2}{*}{\texttt{ATLAS-2108-07586($139fb^{-1}$)}~\cite{ATLAS:2021yqv}}\\{$\tilde{\chi}_2^{0}\tilde{\chi}_1^{\pm}\rightarrow WZ(Wh)\tilde{\chi}_1^0\tilde{\chi}_1^0$}&&\\\\
\multirow{1}{*}{$\tilde{\chi}_{2}^{0}\tilde{\chi}_{1}^{\pm}\rightarrow WZ\tilde{\chi}_{1}^{0}\tilde{\chi}_{1}^{0}$}&\multirow{1}{*}{$2\ell(soft) + jets + \text{E}_\text{T}^{\text{miss}}$}&\texttt{CMS-PAS-SUS-16-025($12.9fb^{-1}$)}~\cite{CMS:2016zvj}\\\\
\multirow{8}{*}{$\tilde{\chi}_{2}^0\tilde{\chi}_1^{\pm}\rightarrow WZ\tilde{\chi}_1^0\tilde{\chi}_1^0$}&\multirow{8}{*}{$n\ell (n\geq2) + nj(n\geq0) + \text{E}_\text{T}^{\text{miss}}$}&\texttt{ATLAS-1712-08119($36.1fb^{-1}$)}~\cite{ATLAS:2017vat}\\&&\texttt{ATLAS-1803-02762($36.1fb^{-1}$)}~\cite{ATLAS:2018ojr}\\&&\texttt{ATLAS-1911-12606($139fb^{-1}$)}~\cite{ATLAS:2019lng}\\&&\texttt{ATLAS-2106-01676($139fb^{-1}$)}~\cite{ATLAS:2021moa}\\&&\texttt{CMS-SUS-16-039($35.9fb^{-1}$)}~\cite{CMS:2017moi}\\&&\texttt{CMS-SUS-16-048($35.9fb^{-1}$)}~\cite{CMS:2018kag}\\&&\texttt{CMS-SUS-20-001($137fb^{-1}$)}~\cite{CMS:2020bfa}\\\\
\multirow{2}{*}{$\tilde{\chi}_{1}^{0}\tilde{\chi}_{1}^{0}\rightarrow ZZ(Zh)(hh)\tilde{G}\tilde{G}$}&\multirow{2}{*}{$n\ell(n\geq2)+ nj(n\geq0) + \text{E}_\text{T}^{\text{miss}}$}&\texttt{CMS-SUS-16-039($35.9fb^{-1}$)}~\cite{CMS:2017moi}\\&&\texttt{CMS-SUS-20-001($137fb^{-1}$)}~\cite{CMS:2020bfa}\\\\
\multirow{2}{*}{$\tilde{\chi}_{2}^0\tilde{\chi}_1^{\pm}\rightarrow Wh\tilde{\chi}_1^0\tilde{\chi}_1^0$}&\multirow{2}{*}{$n\ell(n\geq1) + nb(n\geq0) + nj(n\geq0) + \text{E}_\text{T}^{\text{miss}}$}&\texttt{ATLAS-1909-09226($139fb^{-1}$)}~\cite{ATLAS:2020pgy}\\&&\texttt{CMS-SUS-16-039($35.9fb^{-1}$)}~\cite{CMS:2017moi}\\\\
\multirow{1}{*}{$\tilde{\chi}_{2}^{0}\tilde{\chi}_{1}^{\pm}\rightarrow Wh\tilde{\chi}_{1}^{0}\tilde{\chi}_{1}^{0}$}&\multirow{1}{*}{$2\gamma + nl(n\geq 1) + \text{E}_\text{T}^{\text{miss}}$}&\texttt{ATLAS-2004-10894($139fb^{-1}$)}~\cite{ATLAS:2020qlk}\\\\
\multirow{1}{*}{$\tilde{\chi}_{2}^{0}\tilde{\chi}_{1}^{\pm}\rightarrow Wh\tilde{\chi}_{1}^{0}\tilde{\chi}_{1}^{0}$}&\multirow{1}{*}{$2\gamma + nj(n\geq 2) + \text{E}_\text{T}^{\text{miss}}$}&\texttt{ATLAS-2004-10894($139fb^{-1}$)}~\cite{ATLAS:2020qlk}\\\\
$\tilde{\chi}_2^{0}\tilde{\chi}_1^{\mp}\rightarrow Wh(WZ)\tilde{\chi}_1^0\tilde{\chi}_1^0$&\multirow{2}{*}{$2\gamma + n\ell(n\geq0) + nb(n\geq0) + nj(n\geq0) + \text{E}_\text{T}^{\text{miss}}$}&\multirow{2}{*}{\texttt{ATLAS-1802-03158($36.1fb^{-1}$)}~\cite{ATLAS:2018nud}}\\$\tilde{\chi}_1^{+}\tilde{\chi}_1^{-}\rightarrow WW\tilde{\chi}_1^0\tilde{\chi}_1^0$&&\\\hline
\end{tabular}}
\end{table}

Eq. (\ref{Likelihood}) was used to implement the MultiNest algorithm with ${\it{nlive}} = 8000$~\cite{MultiNest2009} and conduct comprehensive scans over the parameter space shown in Table~\ref{tab:parameter-space}. Samples obtained in this way was then surveyed to see their compatibility with LHC-based electroweakino search results.
This was done by performing a series of Monte Carlo simulations for the following production channels: 
\begin{equation}
\begin{split}
pp &\to \widetilde{\chi}_{\alpha}^{0}\widetilde{\chi}_{\beta}^{0}, \quad \alpha,\beta = 2, 3, 4, 5; \quad \alpha \leq \beta,\\
pp &\to \widetilde{\chi}_{i}^{\pm}\widetilde{\chi}_{j}^{\mp}, \quad  i,j = 1, 2, \\
pp &\to \widetilde{\chi}_{\alpha}^0\widetilde{\chi}_{i}^{\pm}, \quad  \alpha = 2, 3, 4, 5; \quad i = 1, 2. 
\end{split}
\end{equation}
In this process, corresponding cross-sections were first calculated to next-to-leading order (NLO) using program \texttt{Prospino2}~\cite{Beenakker:1996ed}, with LHC fast simulations conducted by the \textsf{SModelS-3.0.0}~\cite{Khosa:2020zar} tool. Specifically, we selected all analyses included in the \textsf{smodels-database-release-3.0.0} and combined signal regions when available, to provide a preliminary assessment. Given the exclusion capabilities of \textsf{SModelS-3.0.0} are limited by the database and strict working prerequisites, we further examined surviving samples by simulating with the Monte-Carlo method the analyses listed in Table~\ref{tab:LHC1}. In more detail, we generated $10^5$ events for these processes using program \texttt{MadGraph\_aMC@NLO}~\cite{Alwall:2011uj, Conte:2012fm}, and implemented parton showers and hadronization using \texttt{PYTHIA8}~\cite{Sjostrand:2014zea}. The resulting event files were then fed into the \texttt{CheckMATE-2.0.37}~\cite{Drees:2013wra,Dercks:2016npn, Kim:2015wza} package, which integrates the modular \texttt{DELPHES-3.5}~\cite{deFavereau:2013fsa} framework for detector simulations. The $R$ value, defined as $R \equiv \max\{S_i/S_{i,obs}^{95}\}$, was subsequently calculated for all involved analyses, where $S_i$ represents a simulated event number for the $i$-th signal region (SR) and $S_{i,obs}^{95}$ is the corresponding $95\%$ C.L. upper limit. A value of $R > 1 $ typically implies the sample is experimentally excluded if the involved uncertainties are neglected~\cite{Cao:2021tuh}, while $R < 1$ indicates consistency with experimental data.

\begin{table}[t]
    \caption{Primary annihilation channels and their proportions for light DM in both $h_1$ and $h_2$ scenarios, with $M_1=150~{\rm GeV}$ and $f=u,d,c,s,b,\tau$. The first row of each scenario takes into account all samples that meet LZ 2024 constraints. The second row corresponds to those further satisfying the LHC constraints, namely predicting an $R$ value less than unity. \label{tab:h1h2-150}}
    \centering
    \vspace{0.3cm}
    \resizebox{0.85\textwidth}{!}{
\begin{tabular}{c|c|c|c|c|c}
\hline\hline
\multicolumn{1}{c|}{\textbf{Scenario}} & \multicolumn{1}{c|}{\textbf{Sample Size}} & {\textbf{$\tilde{\chi}_1^0 \tilde{\chi}_1^0  \to  h_s  h_s$}} & \multicolumn{1}{c|}{\textbf{$\tilde{\chi}_1^0 \tilde{\chi}_1^0  \to  A_s  A_s$}} & \multicolumn{1}{c|}{\textbf{$\tilde{\chi}_1^0 \tilde{\chi}_1^0  \to  h_s  A_s$}} & \multicolumn{1}{c}{\textbf{$\tilde{\chi}_1^0 \tilde{\chi}_1^0  \to  f  \bar f$}} \\ \hline
\multirow{2}{*}{$h_1$} & 8983 & $0$ & $98.3\%$ & $1.0\%$ & $0.6\%$ \\ \cline{2-6}
 & 6379 & $0$ & $98.3\%$ & $1.2\%$ & $0.4\%$ \\ \hline
\multirow{2}{*}{$h_2$} & 6155 & $0$ & $9.8\%$ & $90.0\%$ & $0$ \\ \cline{2-6}
 & 4874 & $0$ & $9.1\%$ & $90.8\%$ & $0$ \\ \hline
\end{tabular}}
\end{table}

\subsection{Numerical results}\label{sec:Nr}

The included parameter space scan produced 23732 samples for the $h_1$ scenario and 21724 samples for the $h_2$ scenario. It is notable that the LZ experiment recently published its 2024 results~\cite{LZ:2024zvo}, prompting us to refine these samples using new constraints that significantly reduced sample sizes to 8983 and 6155, respectively. The previously mentioned LHC constraints further narrowed the surviving samples to 6379 for the $h_1$ 
scenario and 4874 for the $h_2$ scenario. 

To characterize light DM properties comprehensively, Table \ref{tab:h1h2-150} classifies samples satisfying both the LZ 2024 constraints and the LHC threshold criteria $R < 1$ by their dominant annihilation channels. The analysis identifies $\tilde{\chi}_1^0 \tilde{\chi}_1^0 \to A_s A_s$ as the predominant channel in the $h_1$ scenario. This $p$-wave process proceeds under the threshold conditions $|m_{\tilde{\chi}_1^0}| \gtrsim m_{A_s}$ and $m_{h_s} \gtrsim 2 | m_{\tilde{\chi}_1^0} |  - m_{A_s}$. At freeze-out temperature, the thermally-averaged cross-section \(\left\langle \sigma v \right\rangle_{x_F}^{A_s A_s}\) can be approximated by~\cite{Arcadi:2017kky}:
\begin{eqnarray} \label{eq:sigvAsAs2}
\left\langle \sigma v \right\rangle_{x_F}^{A_s A_s} & \simeq & \frac{v^2_F}{128 \pi m_{\tilde{\chi}_1^0}^2} \sqrt{1 - \frac{m_{A_s}^2}{m_{\tilde{\chi}_1^0}^2}}  \nonumber \\
& &\times~ 8\kappa^4\left \{ \frac{1}{3} \frac{ [(\frac{m_{A_s}}{m_{\tilde{\chi}_1^0}})^2 - 1]^2}{[\frac12(\frac{m_{A_s}}{m_{\tilde{\chi}_1^0}})^2 - 1]^4} + \frac{ (\frac{A_\kappa-m_{\tilde S}}{m_{\tilde{\chi}_1^0}})^2  }{[(\frac{m_{h_s}}{m_{\tilde{\chi}_1^0}})^2 - 4]^2+(\frac{m_{h_s}}{m_{\tilde\chi^0_1}})^2(\frac{\Gamma_{h_s}}{m_{\tilde\chi^0_1}})^2}  \right \}, \quad \quad \quad 
\end{eqnarray}
where $v_F \simeq 0.3$ represents the DM velocity at freeze-out. The first and second terms in the braces arise from $t$- and $s$-channel Feynman diagrams, respectively. 
This formula reveals several key features:
\begin{itemize}
\item \textbf{Non-interference property}:  The $s$- and $t$-channel contributions do not interfere. Both scale as $\kappa^4/m_{\tilde{\chi}_1^0}^2$, modulated by a common phase-space factor $\sqrt{1- R_1}$ with $R_1 \equiv (m_{A_s}/m_{\tilde{\chi}_1^0})^2$.    
     
\item \textbf{$t$-channel behavior}: For fixed $m_{\tilde{\chi}_1^0}$, the $t$-channel contribution depends solely on $\kappa$ and the squared mass ratio $R_1$. When normalized to its value at $R_1 =0$,  it equals approximately $0.87$, $0.68$, $0.42$, and $0.14$ for $R_1 = 0.2$, $0.4$, $0.6$, and $0.8$, respectively — decreasing monotonically as $R_1$ grows. Thus, its overall contribution is more sensitive to $\kappa$ than to the squared mass ratio $R_1$.   
    
\item \textbf{$s$-channel complexity}: In contrast to the $t$-channel, the $s$-channel contribution depends on additional parameters: the mass of $h_s$, its width $\Gamma_{h_s}$, and the combination $D \equiv (A_\kappa - m_{\tilde S})/m_{\tilde{\chi}_1^0}$ arising from the $h_s A_s A_s$ coupling. Among the samples from our scan, $\Gamma_{h_s}$ remains very small and $h_s$ rarely approaches resonance during annihilation. Consequently, the width term can be safely neglected. 

\item \textbf{Magnitude comparison}: While the $t$‑channel amplitude in braces never exceeds $1/3$, the $s$‑channel component can reach several orders of magnitude larger when $|D|$ is substantial and $m_{h_s}$ approaches $2 |m_{\tilde{\chi}_1^0}|$. Consequently, the $s$‑channel annihilation often governs the relic density.
\end{itemize}

\begin{figure}[t]
\centering 
{\tiny
\includegraphics[scale=0.5]{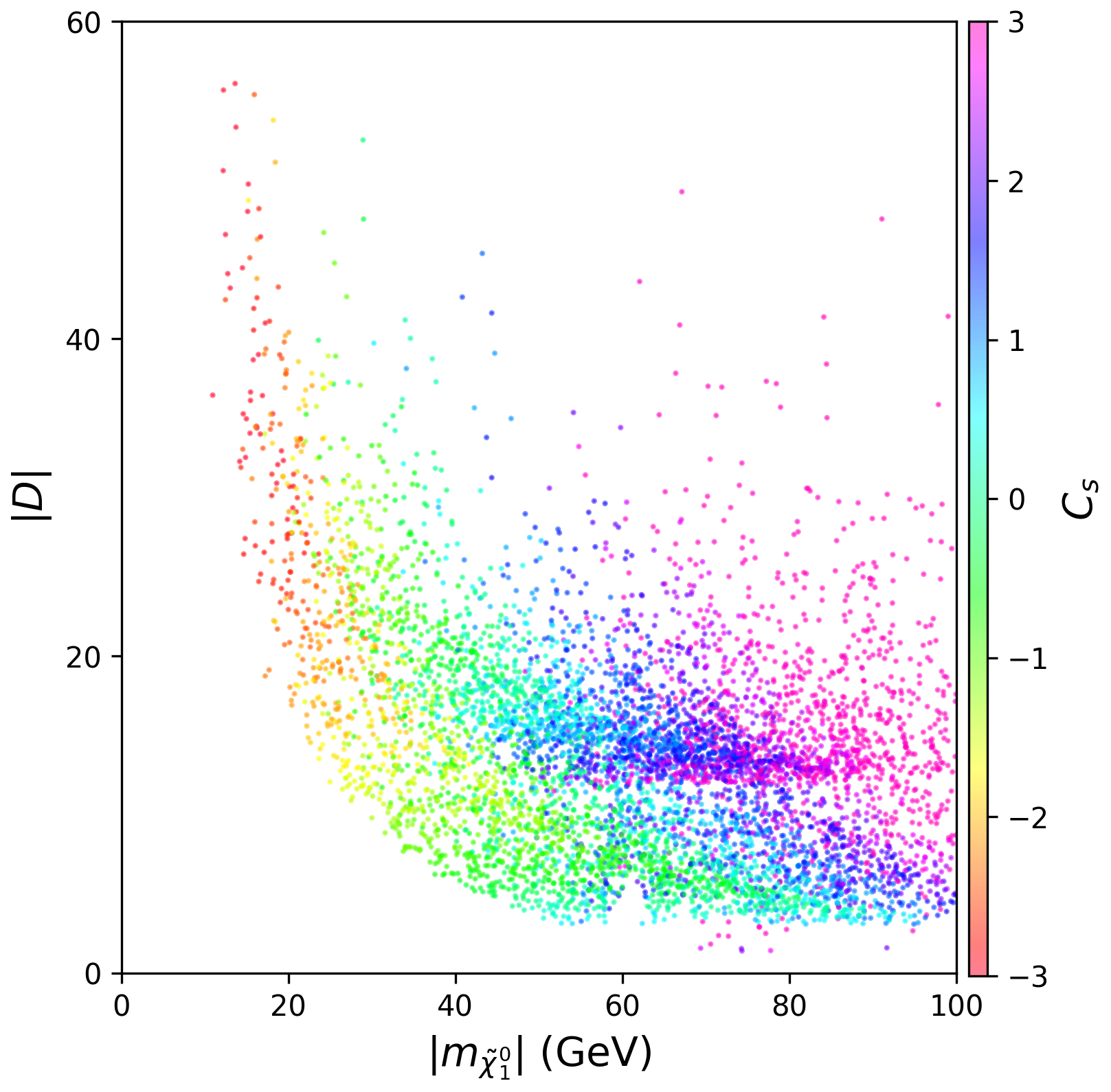} \quad\quad
\includegraphics[scale=0.5]{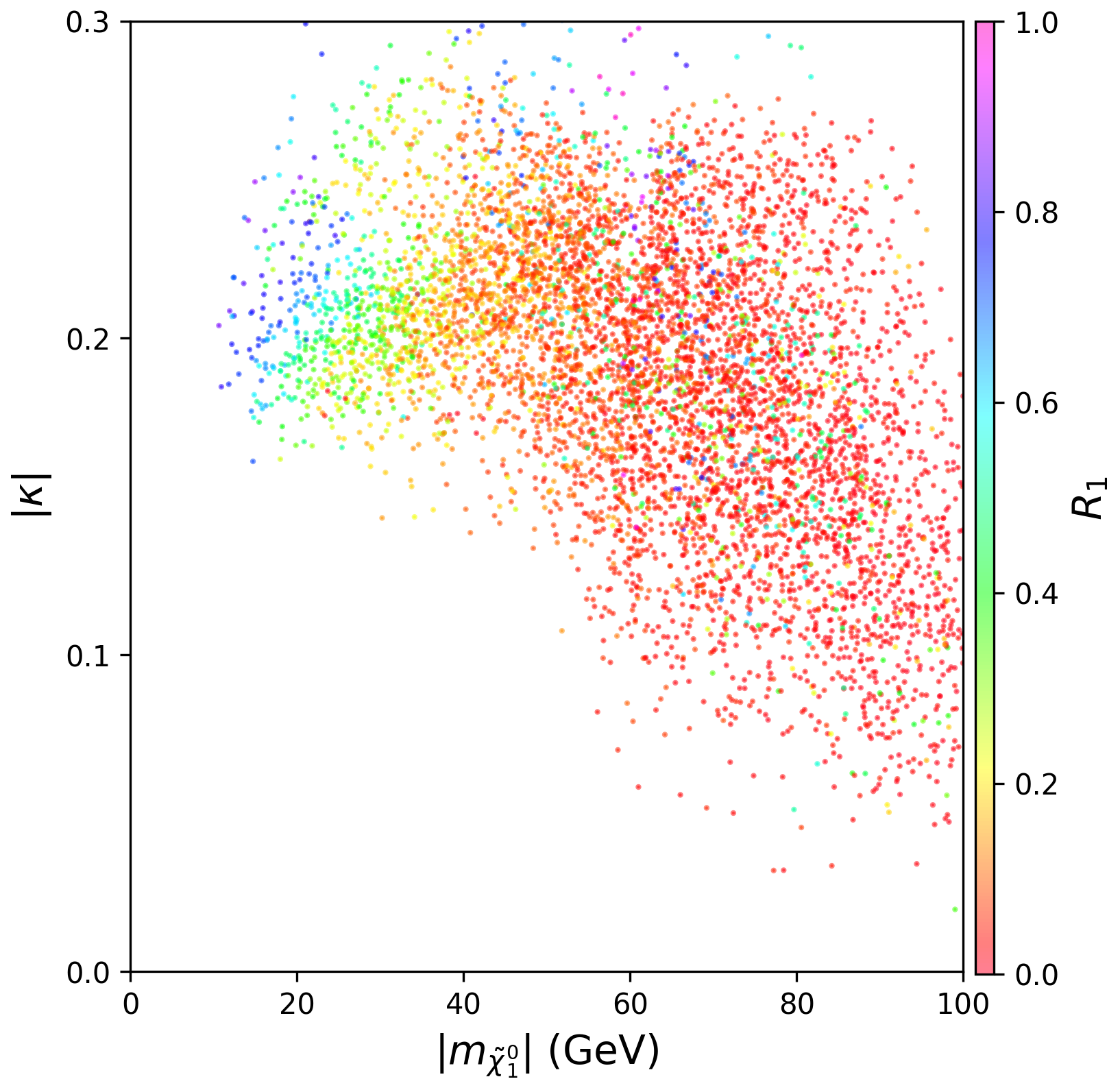}
\\[-0.4cm]
}
\caption{Scatter plots for samples dominated by $\tilde{\chi}_1^0 \tilde{\chi}_1^0 \to A_s A_s$ annihilation and satisfying all experimental constraints in the $h_1$ scenario. Left panel:  $|D| \equiv |(A_\kappa - m_{\tilde S})/m_{\tilde{\chi}_1^0}|$ versus $|m_{\tilde{\chi}_1^0}|$, with color scale $C_s$ defined by $2\{ \ln |(A_\kappa-m_{\tilde S})/m_{\tilde\chi^0_1} | -\ln |(m_{h_s}/m_{\tilde \chi^0_1})^2-4)| \}$, indicating relative $s$-channel strength in Eq.~(\ref{eq:sigvAsAs2}). Right panel: $|\kappa|$ versus $|m_{\tilde{\chi}_1^0}|$, colored by the squared mass ratio $R_1 \equiv (m_{A_s}/m_{\tilde{\chi}_1^0})^2$.  \label{fig:ann-h1-150}}
\end{figure}

Figure~\ref{fig:ann-h1-150} projects the samples dominated by $\tilde{\chi}_1^0 \tilde{\chi}_1^0 \to A_s A_s$ annihilation onto the  $|m_{\tilde{\chi}_1^0}|$–$|D|$ and $|m_{\tilde{\chi}_1^0}|$–$|\kappa|$ planes. Compared to the full parameter space defined in Table~\ref{tab:parameter-space}, accommodating the observed relic density while satisfying the LZ 2024 constraints requires $|\kappa|\lesssim 0.3$ and $|m_{\tilde{\chi}_1^0}| \gtrsim 10~{\rm GeV}$. Further analysis reveals distinct mass-dependent behaviors across three regimes:

\begin{itemize}
\item \textbf{Light DM regime ($10-30~{\rm GeV}$):} As shown in the left panel of Fig.~\ref{fig:ann-h1-150}, the $s$-channel contribution remains subdominant due to $m_{h_s}^2 \gg m_{\tilde{\chi}_1^0}^2$,  given that $m_{h_s}$ exceeds $125~{\rm GeV}$.  It gradually strengthens and becomes comparable with the $t$-channel as $|m_{\tilde{\chi}_1^0}|$  approaches $30~{\rm GeV}$. 
By contrast, despite substantial $R_1$ values (numerically $0.6 \lesssim R_1 \lesssim 0.8$) that suppress the $t$-channel amplitude, sufficiently large $\kappa$ ensures $t$-channel dominance in determining the relic abundance, as illustrated in the right panel of Fig.~\ref{fig:ann-h1-150}. 
\item \textbf{Intermediate masses ($30-60~{\rm GeV}$):} The $s$-channel contribution increases progressively and begins to surpass the $t$-channel, which diminishes due to its inverse proportionality to $m_{\tilde{\chi}_1^0}^2$.
\item \textbf{Heavy DM regime ($> 60~{\rm GeV}$):}  A dramatic transition occurs wherein the $s$-channel typically dominates and intensifies further with increasing mass. This dominance arises through two mechanisms. 
First, most samples satisfy $0.6\lesssim m_{h_s}/(2|m_{\tilde\chi^0_1}|)\lesssim 1.4$, where near-resonance effects enhance the $s$-channel. Second, for samples far from resonance,  $|A_\kappa/m_{\tilde{\chi}_1^0}|$ becomes sufficiently large to ensure $s$-channel dominance. These mechanisms allow $\kappa$ as low as $0.04$ to produce the measured abundance at $|m_{\tilde{\chi}_1^0}| \simeq 100~{\rm GeV}$. 
\end{itemize}

In the $h_2$ scenario, DM achieves the observed relic abundance primarily through the $s$-wave annihilation channel $\tilde{\chi}_1^0 \tilde{\chi}_1^0 \to h_s A_s$, as listed in Table~\ref{tab:h1h2-150}. This process requires the kinematic condition $2 |m_{\tilde{\chi}_1^0}| \gtrsim m_{h_s} + m_{A_s}$, suggesting that all the singlet-dominated particles are likely to be simultaneously light. Compared with the $h_1$ scenario with a massive $h_s$, 
the $h_2$ case is more tightly constrained by current Higgs coupling measurements and by LHC searches for exotic decays of the SM‑like Higgs boson~\cite{Heng:2025wng}. The underlying reason is that, according to the approximation in Eq.~(\ref{Approximations}), $h$ typically contains a more substantial singlet admixture in this scenario. 
Consequently, its couplings to singlet‑dominated states become stronger when $\kappa$ is sizable, leading to potentially large branching fractions for exotic decays such as 
$h \to \tilde{\chi}_1^0 \tilde{\chi}_1^0, h_s h_s$, and $A_s A_s$. 

\begin{figure}[t]
\centering
{\tiny
\includegraphics[scale=0.5]{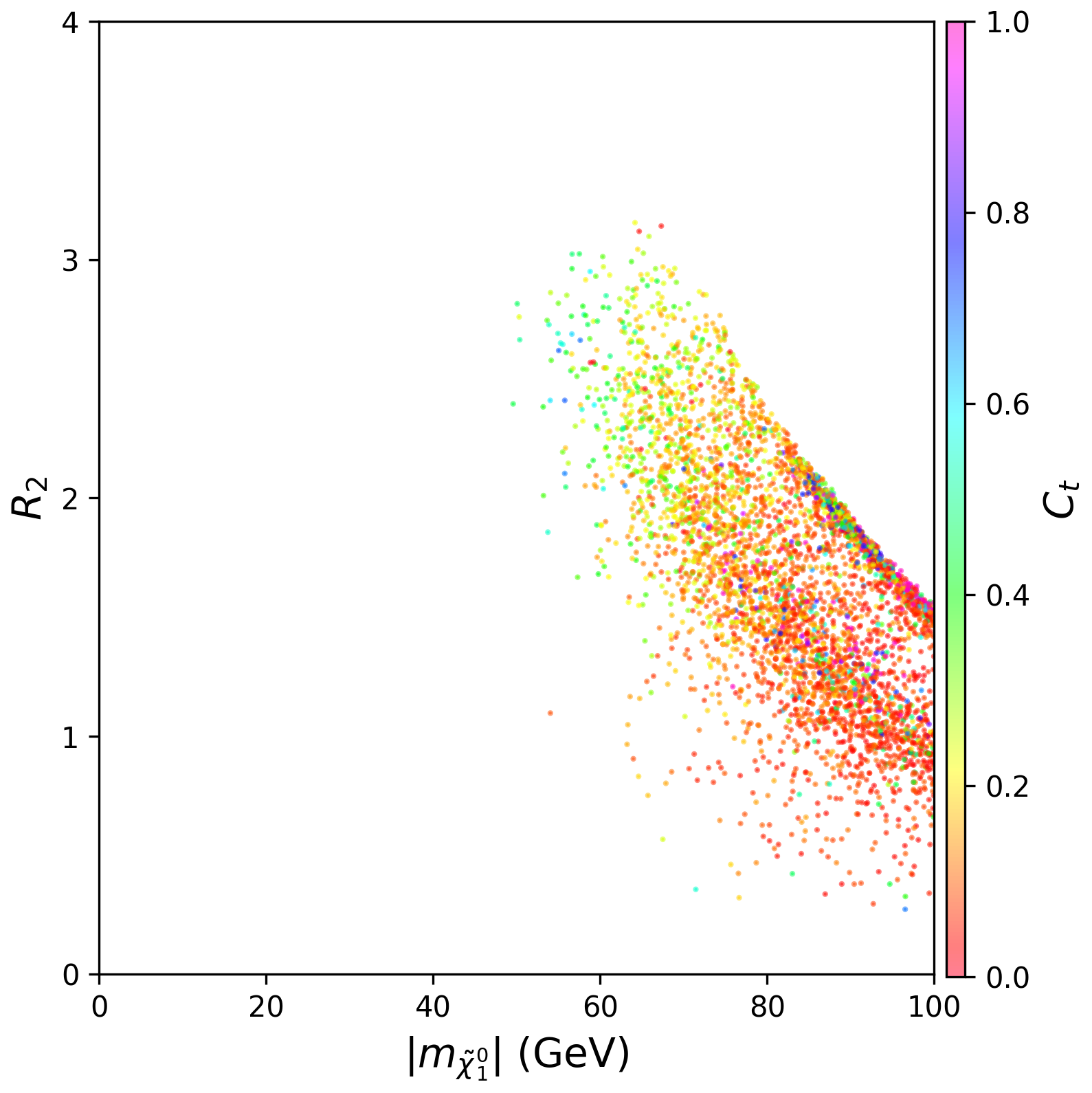} \quad\quad
\includegraphics[scale=0.5]{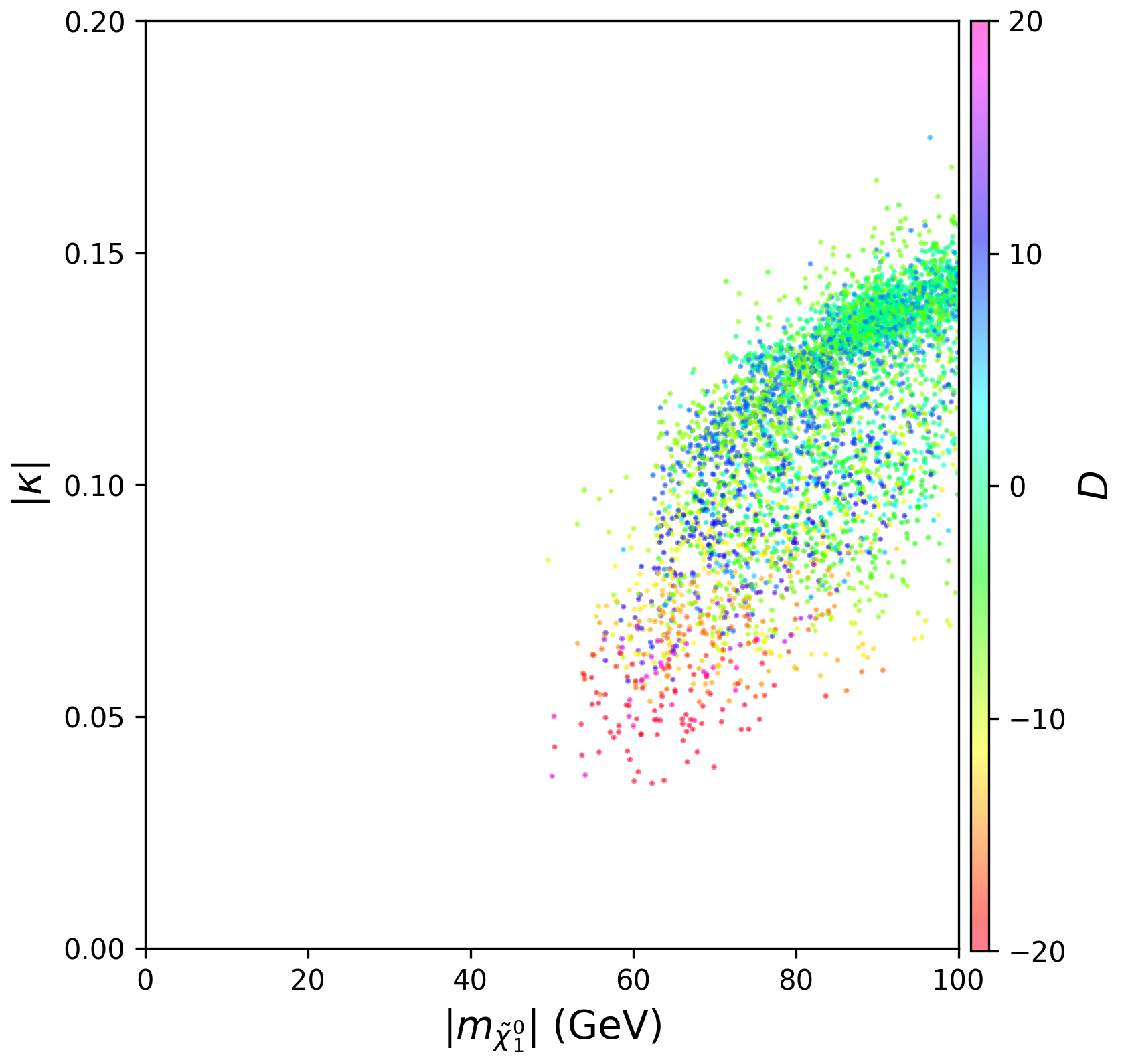}
\\[-0.4cm]
}
\caption{ Scatter plots for samples dominated by $\tilde{\chi}_1^0 \tilde{\chi}_1^0 \to h_s A_s$ annihilation and satisfying all experimental constraints in the $h_2$ scenario. Left panel: $R_2 \equiv (m_{h_s}/m_{\tilde{\chi}_1^0})^2$ versus $|m_{\tilde{\chi}_1^0}|$, with a color bar $C_t$ defined by $8 m_{A_s}^2/(4 m_{\tilde \chi^0_1}^2- m_{h_s}^2)$, which measures the partial $t$-channel contributions of Eq.~(\ref{eq:sigvPhiPhi1}). Right panel:  $|\kappa|$ versus $|m_{\tilde{\chi}_1^0}|$, with a color bar representing the 
$s$-channel contribution $D\equiv (A_\kappa - m_{\tilde S})/m_{\tilde{\chi}_1^0}$ in Eq.~(\ref{eq:sigvPhiPhi1}). \label{fig:ann-h2-150}}
\end{figure}

The combined constraints from Higgs physics and the measured relic density yield a preferred particle spectrum characterized by $m_{h_s} > |m_{\tilde{\chi}_1^0}| \gtrsim 60~{\rm GeV}$ and $m_{A_s}^2/m_{\tilde{\chi}_1^0}^2 < 1/4 $  for most samples, with an absolute upper limit of  $m_{A_s}^2/m_{\tilde{\chi}_1^0}^2 < 1/2 $ across all surviving samples.
These features are illustrated in the left panel of Fig.~\ref{fig:ann-h2-150}, where samples dominated by the $\tilde{\chi}_1^0 \tilde{\chi}_1^0 \to h_s A_s$ annihilation channel are projected onto  $R_2 \equiv m_{h_s}^2/m_{\tilde{\chi}_1^0}^2$ versus  $|m_{\tilde{\chi}_1^0}|$ plane. The color scale $C_t \equiv 8 m_{A_s}^2/(4 m^2_{\tilde{\chi}_1^0} - m^2_{h_s})$ quantifies the partial $t$-channel contribution as introduced below. The right panel displays the same samples in $|\kappa|$ versus  $|m_{\tilde{\chi}_1^0}|$ plane, where the color variable $D \equiv (A_\kappa - m_{\tilde S})/m_{\tilde{\chi}_1^0}$ represents the combination governing the $s$‑channel contribution.

The thermally-averaged cross-section at freeze-out for this channel, $\left\langle \sigma v \right\rangle_{x_F}^{h_s A_s}$, is expressed as~\cite{Meng:2024lmi,Baum:2017enm,Griest:1990kh}:
\begin{eqnarray} \label{eq:sigvPhiPhi}
\left\langle \sigma v \right\rangle_{x_F}^{h_s A_s} \simeq && \frac{1}{256 \pi m_{\tilde{\chi}_1^0}^2} \left\{ \left[4 -\frac{\left(m_{h_s} + m_{A_s}\right)^2}{m_{\tilde{\chi}_1^0}^2}\right] \left[4-\frac{\left(m_{h_s} - m_{A_s}\right)^2}{m_{\tilde{\chi}_1^0}^2}\right] \right\}^{1/2} | {\cal{A}}_s + {\cal{A}}_t |^2, \nonumber
\end{eqnarray}
where the $s$- and $t$-channel amplitudes, ${\cal{A}}_s$ and ${\cal{A}}_t$,  depends primarily on the spectrum and couplings of the singlet-dominated fields. For small $\lambda$ and $m_{A_s}^2 \ll 4 m_{\tilde{\chi}_1^0}^2$, this expression simplifies to:
\begin{eqnarray}\label{eq:sigvPhiPhi1}
\left\langle \sigma v \right\rangle_{x_F}^{h_s A_s}  &\simeq &\frac{1}{256 \pi m_{\tilde{\chi}_1^0}^2}\left[4-(\frac{m_{h_s}}{m_{\tilde{\chi}_1^0}})^2\right] \left|{\cal{A}}_s+{\cal{A}}_t\right|^2,
\end{eqnarray}
with the amplitudes approximated as:
\begin{eqnarray}\label{eq:sigvPhiPhi2}
{\cal{A}}_s &\simeq &  \kappa^2  D, \quad \quad
{\cal{A}}_t \simeq  \kappa^2 ( 4 + C_t ).
\end{eqnarray}
The color coding in Fig. \ref{fig:ann-h2-150} reveals that the amplitude $\mathcal{A}_s$ varies substantially from $-20\kappa^2$ to $20\kappa^2$, while $\mathcal{A}_t$ 
remains relatively stable between $4 \kappa^2$ to $5 \kappa^2$. The sign of  $\mathcal{A}_s$ determines whether it interferes constructively or destructively with $\mathcal{A}_t$, thereby significantly affecting the required magnitude of $\kappa$ for achieving the measured relic density. Specifically, when $D \simeq 20$, the $s$-channel contribution dominates and constructive interference allows $|\kappa|$ as low as $0.03$ for $|m_{\tilde{\chi}_1^0}| \simeq 100~{\rm GeV}$ to acquire the density.  Conversely, in the $|D| \lesssim 10$ regime — which characterizes the majority of surviving samples — the two amplitudes become comparable. When destructive interference occurs, larger $|\kappa|$ values up to $0.16$ are necessary at  $|m_{\tilde\chi^0_1}|\simeq 100~{\rm GeV}$ to compensate for the reduced total amplitude. These relationships are evident in the right panel of Fig.~\ref{fig:ann-h2-150}.

We clarify that in the $h_2$ scenario, the singlet-like Higgs mass satisfies $m_{h_s} \lesssim 125~{\rm GeV}$ and this upper limit is consistently achieved when $|m_{\tilde{\chi}_1^0}| \gtrsim 70~{\rm GeV}$ for the obtained samples. 
Consequently, the maximum reach of $R_2$ decreases monotonically with increasing $|m_{\tilde{\chi}_1^0}|$, as seen in the left panel of Fig.~\ref{fig:ann-h2-150}. Besides, since $\left\langle \sigma v \right\rangle_{x_F}^{h_s A_s}$ is proportional to $\kappa^4/m_{\tilde{\chi}_1^0}^2$, larger $|\kappa|$ values are required to maintain the measured relic abundance as $|m_{\tilde{\chi}_1^0}|$ increases,  a trend clearly visible in the right panel.  Additionally, as exhibited by the correlation between the two panels, for fixed $|m_{\tilde{\chi}_1^0}| \simeq 60~{\rm GeV}$, an increase in $R_2$ — arising from enhanced $m_{h_s}$ and consequently reduced annihilation phase space — requires a correspondingly larger $\kappa^2 D$ to preserve the measured relic density. This reflects the interplay of the parameters.

Next, we turn to DM–nucleon scattering. For $m_{h_s} \lesssim 250~{\rm GeV}$ and $m_H \simeq 2~{\rm TeV}$, the spin-independent cross-section $\sigma^{\rm SI}_{\tilde\chi^0_1,N}$ is dominated by $t$-channel exchange of $h$ and $h_s$, with the leading contribution scaling as $\lambda^2\kappa^2$ due to the presence of a relatively light $h_s$~\cite{Meng:2024lmi}. Crucially, the two exchange amplitudes can interfere destructively, producing a notable cancellation that intricately interplay among samples with different $\lambda \kappa$ values. Furthermore, with $M_1 = 150~{\rm GeV}$ in our analysis, the DM retains a Bino admixture up to $0.1\%$ level. 
Despite being small, this component yields a non-negligible contribution for  $\sigma^{\rm SI}_{\tilde\chi^0_1,N}$ because the Bino-Bino-SM Higgs coupling $C_{\tilde{B} \tilde{B} H_{\rm SM}}$ is much larger than the Singlino-Singlino-SM Higgs coupling $C_{\tilde{S} \tilde{S} H_{\rm SM}}$ when the Higgsino mass lies around several hundred GeV. 
Additionally, the small $\tilde{H}_d$ component in $\tilde{\chi}_1^0$ may also be non-negligible in the scattering since $\tilde{S}-\tilde{H}_d-H_{\rm SM}$ coupling $C_{\tilde{S} \tilde{H}_d H_{\rm SM}}$ is determined by $\lambda$. These effects, not previously emphasized in literature, can substantially enhance the cancellations among different contributions to $\sigma^{\rm SI}_{\tilde\chi^0_1,N}$. These observations have important implications for satisfying LZ 2024 constraints: strong destructive interference permits larger $\lambda |\kappa|$ values, while weak cancellation favors smaller $\lambda |\kappa|$.

\begin{figure}[t]
\centering
{\tiny
\includegraphics[scale=0.5]{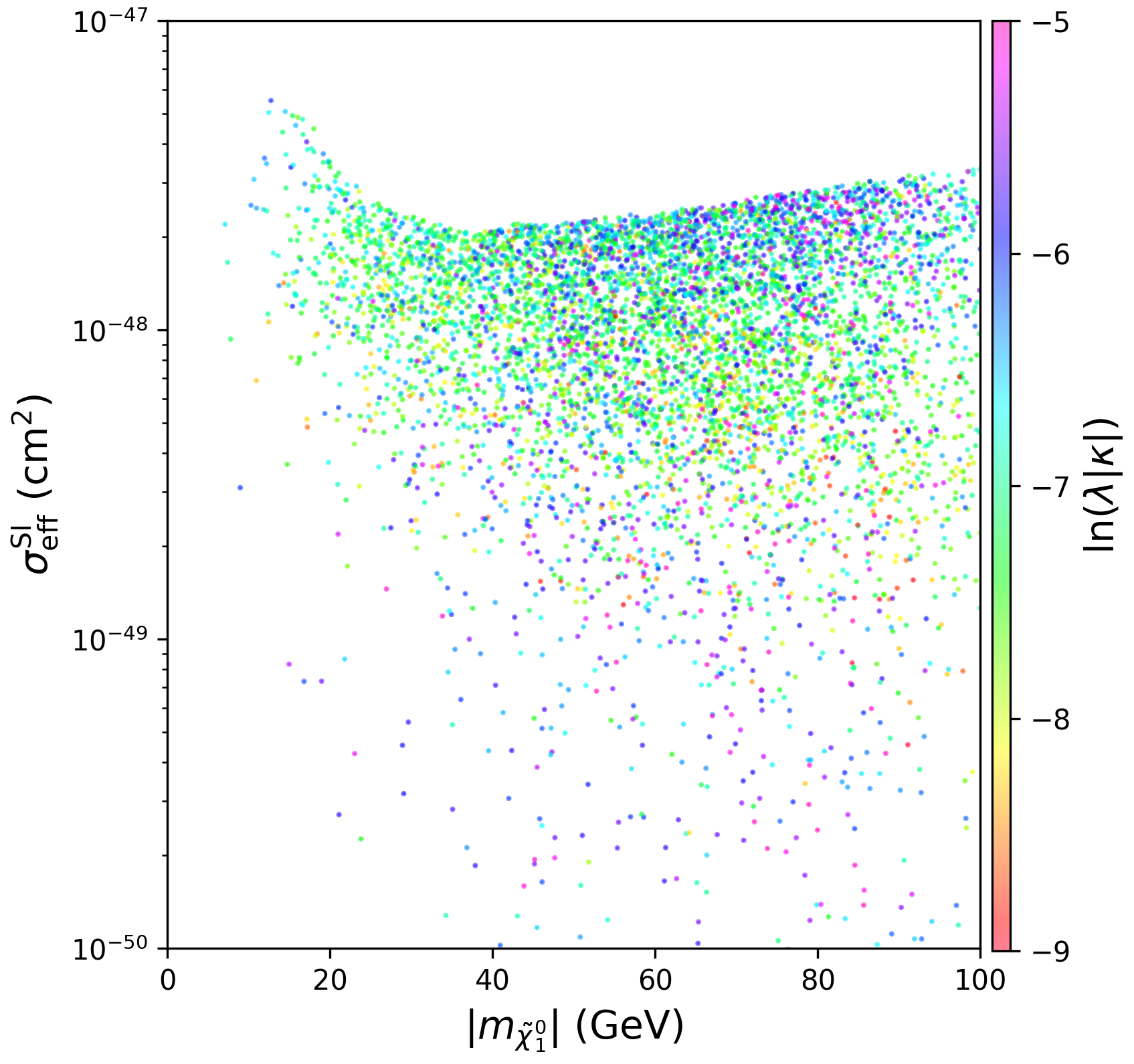} \quad\quad
\includegraphics[scale=0.5]{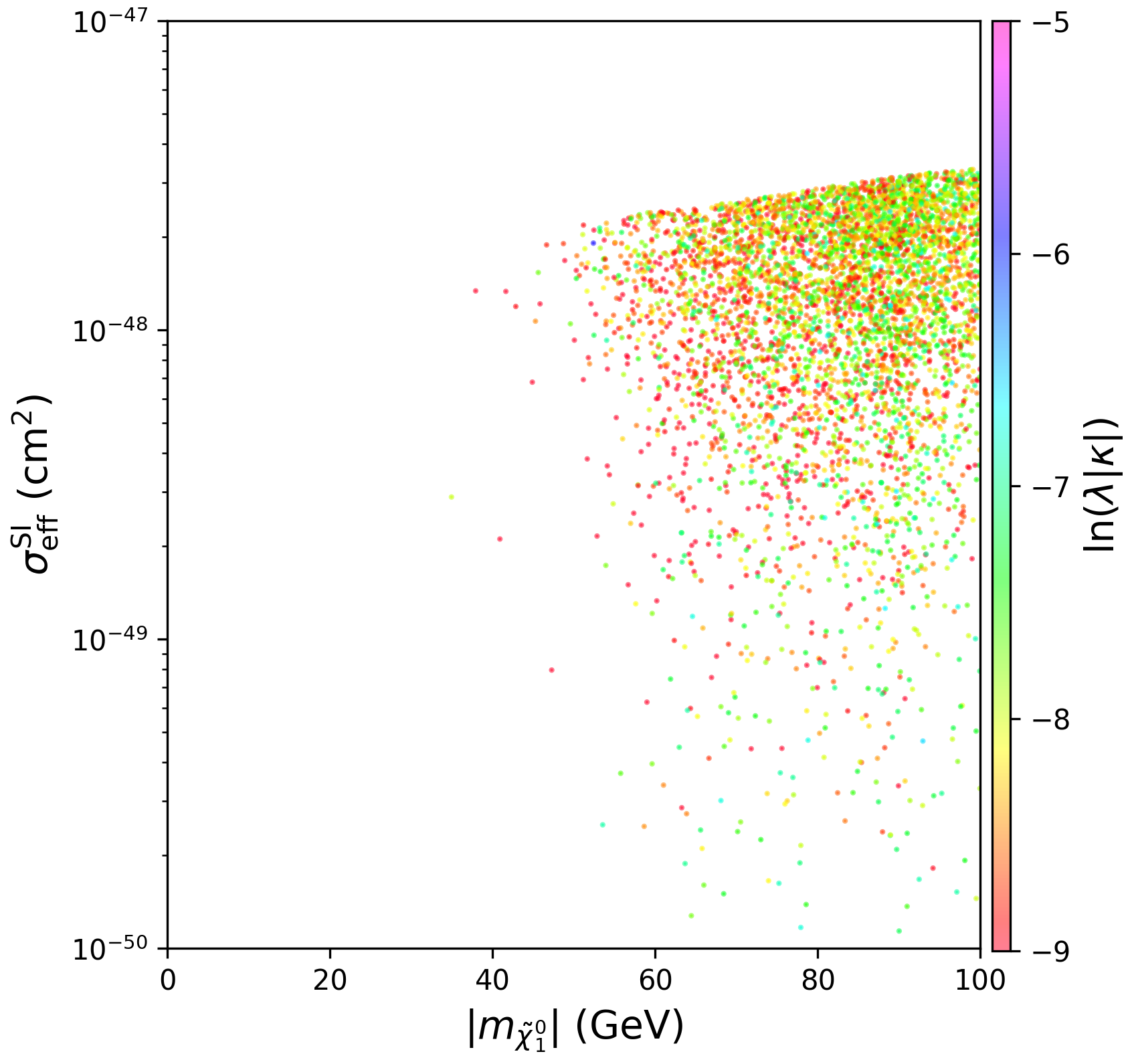} \\[-0.4cm]
}
\caption{Scatter plots of $\sigma_{\rm eff}^{\rm SI}$ versus $|m_{\tilde\chi^0_1}|$ for the samples in the $h_1$ scenario (left panel) and the $h_2$ scenario (right panel). The color bar is defined as $\ln (\lambda |\kappa|)$, describing the coupling strength  involved in the scattering. \label{fig:150-sigmaSI-DM-lambda-kappa}}
\end{figure}

This phenomenon is numerically verified in Fig.~\ref{fig:150-sigmaSI-DM-lambda-kappa}, where the left and right panels display results for the $h_1$ and $h_2$ scenarios, respectively. The figure reveals that the cancellation in the $h_2$ scenario is significantly weaker than in the $h_1$ scenario,  thereby favoring smaller $\lambda |\kappa|$. Nevertheless, despite this difference in cancellation strengths, both scenarios can suppress $\sigma^{\rm SI}_{\tilde\chi^0_1,N}$ to values as low as $10^{-50}~{\rm cm}^2$. Notably, the required smallness of $\lambda$ naturally aligns with constraints from Higgs data fitting, as elaborated in Section~\ref{sec:z3}.

\begin{figure}[t]
\centering
{\tiny
\includegraphics[scale=0.5]{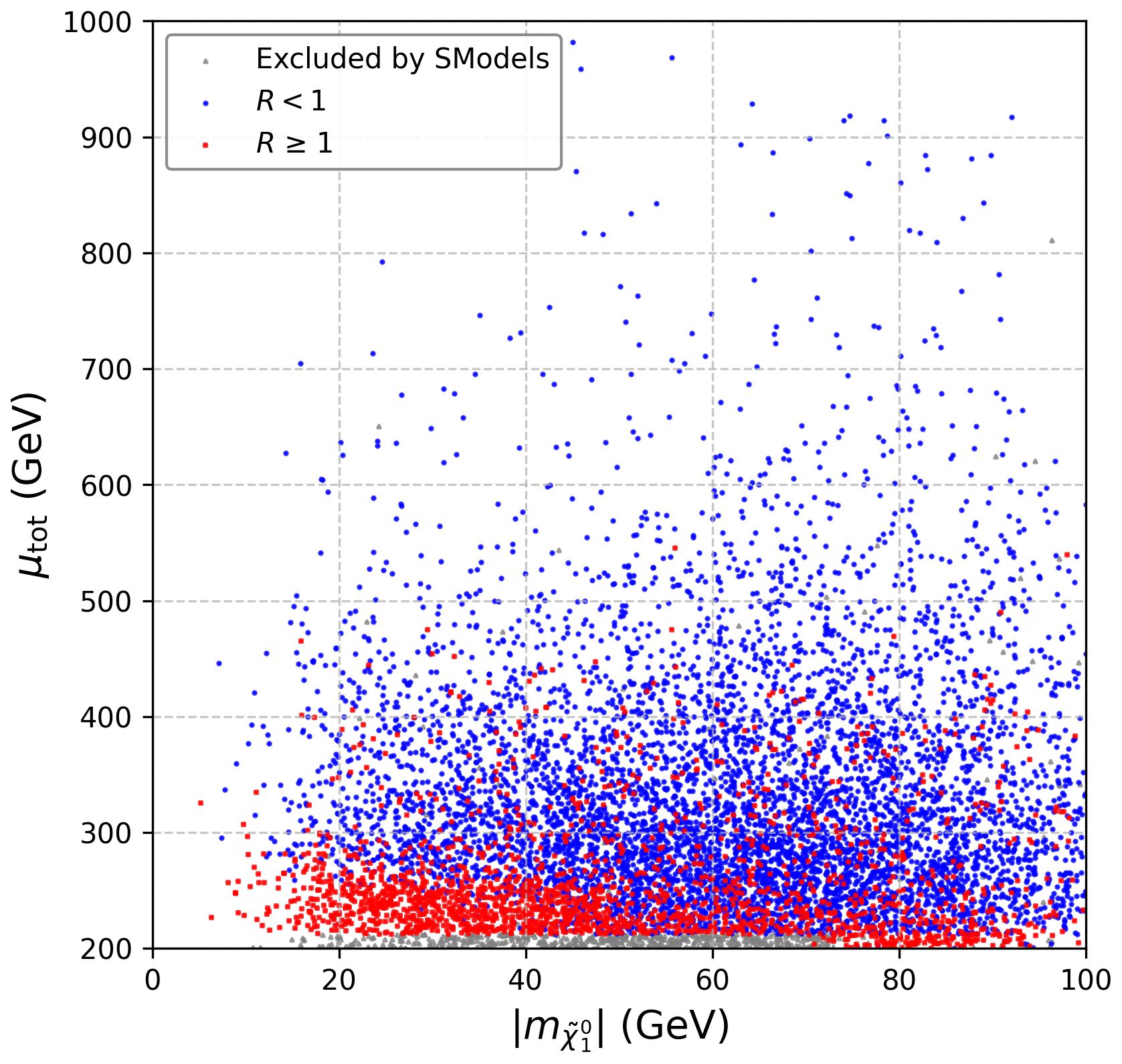} \quad\quad\quad
\includegraphics[scale=0.5]{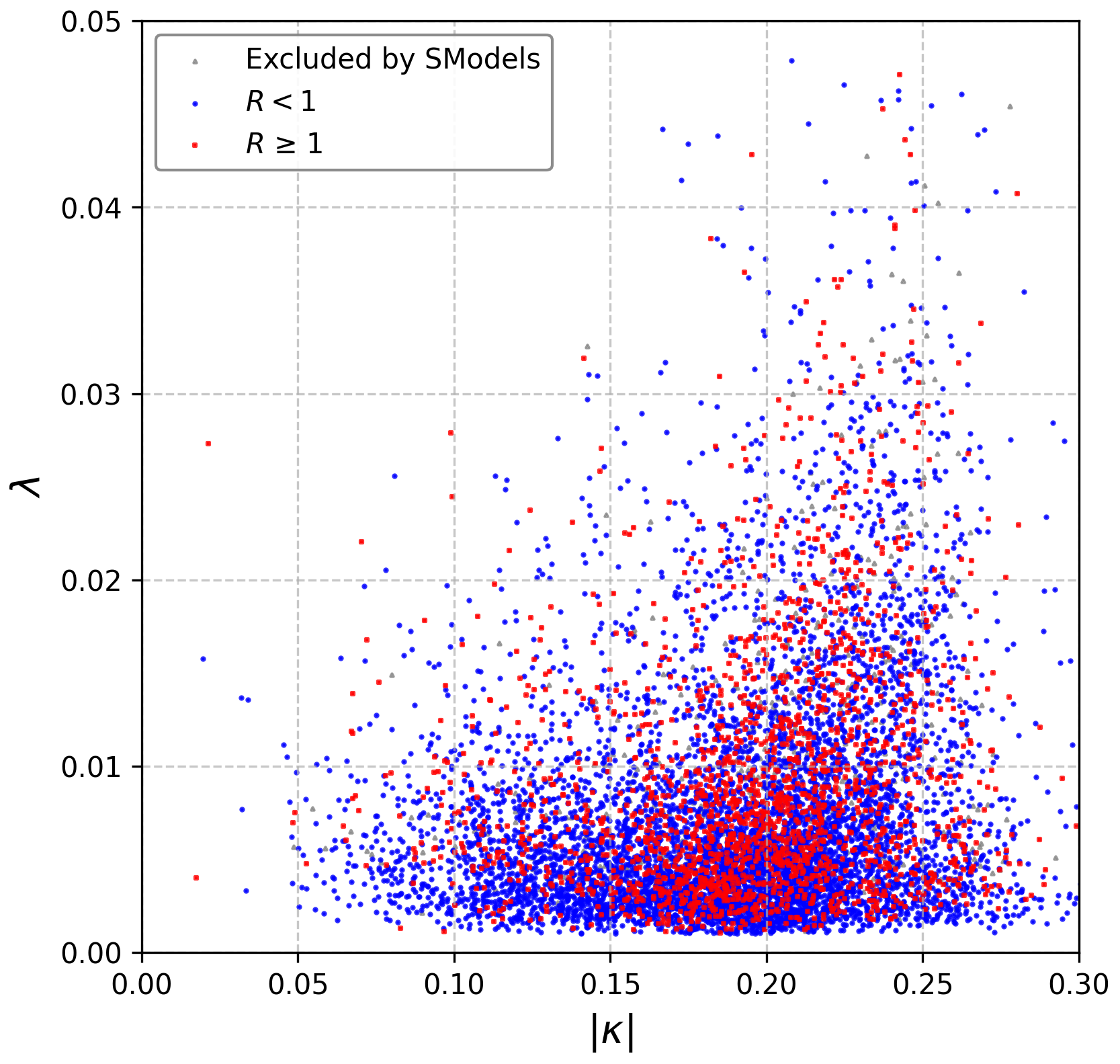} \\[-0.4cm]
\includegraphics[scale=0.5]{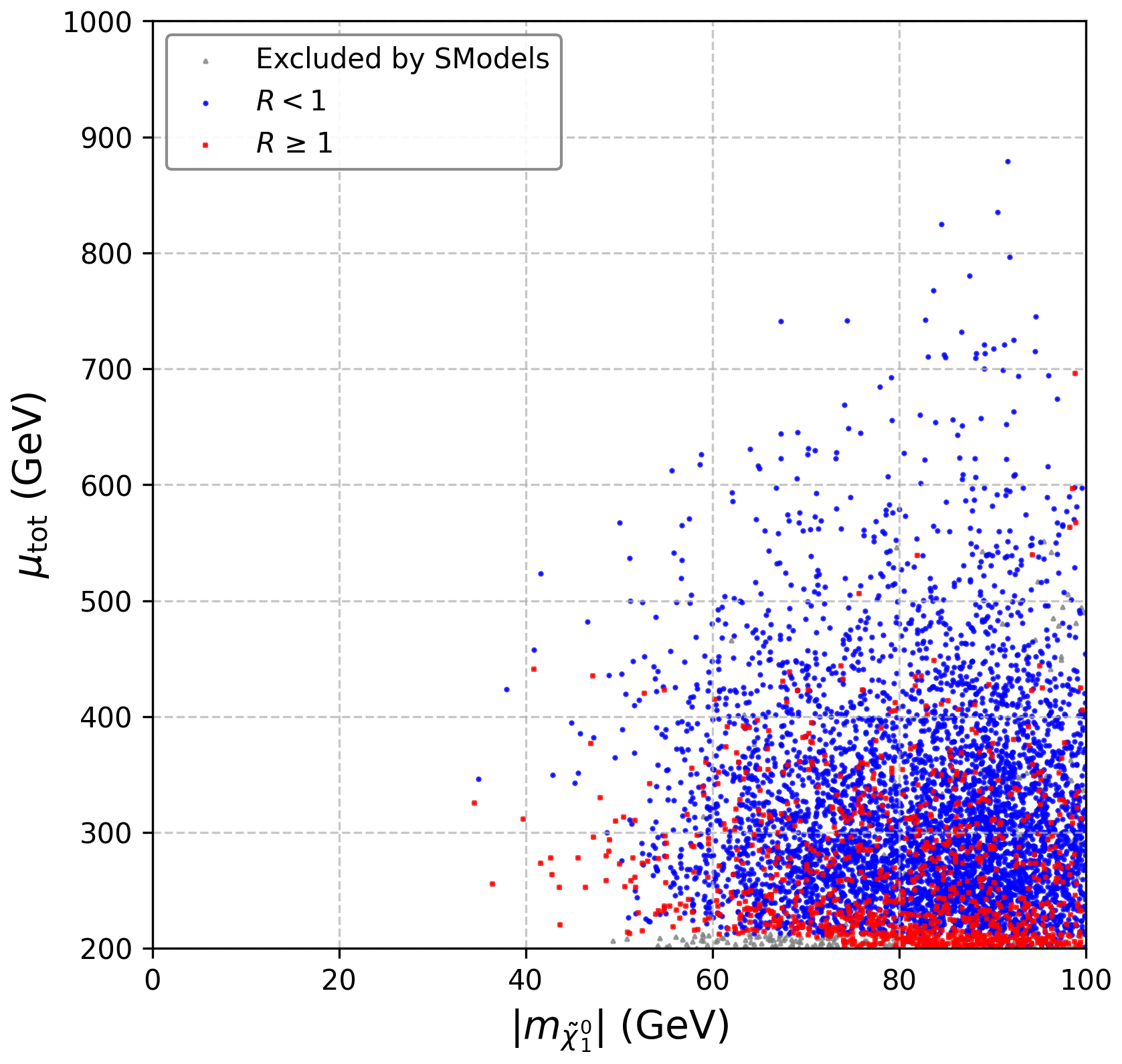} \quad\quad\quad
\includegraphics[scale=0.5]{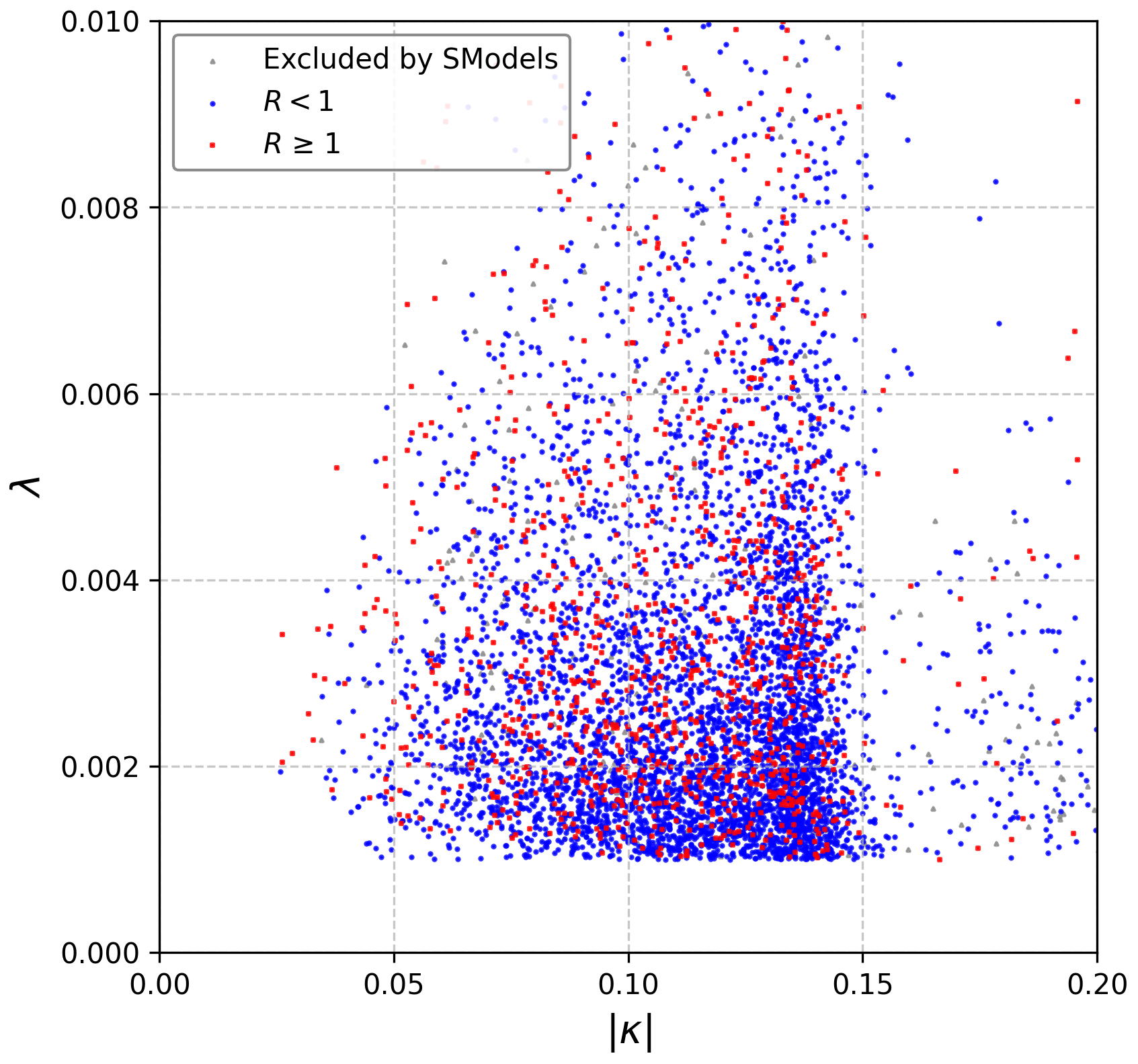} \\[-0.4cm]
}
\caption{LHC constraints on the parameters $|m_{\tilde{\chi}_1^0}|$, $\mu_{\rm tot}$, $|\kappa|$, and $\lambda$. Samples include those obtained by the scans and refined by the LZ 2024 constraints. They are classified by exclusion levels, as depicted in the legend. Gray markers are excluded by \textsf{SModelS-3.0.0}, blue markers ($R < 1$) are consistent with all experiments, and red markers ($R \ge 1$) are excluded by LHC data. The upper panels correspond to the $h_1$ scenario, while the lower panels refer to the $h_2$ scenario.\label{fig:LHC-h1h2-150}}
\end{figure}

Finally, we examine the LHC constraints on supersymmetry. For this purpose, we project samples obtained from parameter scans and refined by LZ 2024 constraints onto $|m_{\tilde{\chi}_1^0}|-\mu_{\rm tot}$ and $|\kappa|-\lambda$ planes, as shown in Fig.~\ref{fig:LHC-h1h2-150}.
Several key observations emerge:
\begin{itemize}
\item When the Bino-like neutralino serves as the next-to-lightest supersymmetric particle (NLSP), the decay cascades are longer and the final states have higher multiplicities (often including a light $A_s$), which degrades search sensitivity and relaxes the LHC bounds.  In this hierarchy, i.e., $\tilde{S} < \tilde{B} < \tilde{H}$, samples with $\mu_{\rm tot}$ as low as $\sim 200~{\rm GeV}$ remain viable, alleviating the fine-tuning in predicting $Z$-boson mass and preserving theoretical naturalness.
\item Samples excluded by our recast are primarily characterized by final states containing three or more leptons, which fall into two categories~\cite{CMS:2017moi}: (i) No $\tau$ leptons, at least two opposite-sign same-flavor (OSSF) lepton pairs (electrons or muons), and missing transverse momentum $p_T^{\rm miss} \in (100,150)~{\rm GeV}$ or $p_T^{\rm miss} \ge 200~{\rm GeV}$. (ii) Two $\tau$ leptons, fewer than two OSSF pairs, and $p_T^{\rm miss} \ge 150~{\rm GeV}$. Note that the substantial missing transverse momentum in these signatures constitutes a characteristic feature of supersymmetric particle production.  

\item The LHC and LZ 2024 constraints exhibit mutual consistency, as surviving samples display stable distributions in $\lambda$ and $|\kappa|$ parameter space. Specifically, in the $h_1$ scenario, $\lambda$ was bounded above by $\lambda \lesssim 0.048$, while $|\kappa|$ could extends up to 0.3. By contrast, the $h_2$ scenario imposes more stringent constraints: $\lambda \leq 0.01$ and $|\kappa| \in (0.03, 0.2)$, with most samples concentrated near $|\kappa| \simeq 0.15$.
\end{itemize}

\begin{table}[t]
\caption{Detailed information for two benchmark points aligning with all currently available experimental results, with $M_1=150~{\rm GeV}$. \label{tab:BP-150}}\
\centering
\resizebox{1\textwidth}{!}
{
\begin{tabular}{crcr|crcr}
\hline \hline
\multicolumn{4}{c|}{\bf Benchmark Point BP1~~~($h_1$ scenario)}                                                                                                & \multicolumn{4}{c}{\bf Benchmark Point BP2~~~($h_2$ scenario)}                                                                                                \\ \hline
$\lambda$             & 0.008& $m_{h_s}$                & 148.6~GeV& $\lambda$             & 0.002& $m_{h_s}$                & 100.4~GeV\\
$\kappa$              & -0.194& $m_{h}$                & 125.4~GeV& $\kappa$              & 0.110& $m_{h}$               & 125.5~GeV\\
$\delta$    & 0.375& $m_{H}$ & 1877.6~GeV& $\delta$    & 0.721& $m_{H}$ & 1840.6~GeV\\
$\tan{\beta}$         & 22.57& $m_{A_s}$                  & 11.7~GeV& $\tan{\beta}$         & 23.88& $m_{A_s}$                  & 10.7~GeV\\
$\mu$                 & 224.9~GeV& $m_{\tilde{\chi}_1^0}$                  & -52.4~GeV& $\mu$                 & 299.0~GeV& $m_{\tilde{\chi}_1^0}$                  & -73.3~GeV\\
$\mu_{\rm eff}$                 & 3.2~GeV& $m_{\tilde{\chi}_2^0}$                  & 141.7~GeV& $\mu_{\rm eff}$                 & 1.0~GeV& $m_{\tilde{\chi}_2^0}$                  & 146.3~GeV\\
$\mu_{\rm tot}$ & 228.1~GeV& $m_{\tilde{\chi}_3^0}$                & -239.0~GeV& $\mu_{\rm tot}$ & 300.0~GeV& $m_{\tilde{\chi}_3^0}$                & -312.1~GeV\\
$A_\lambda$            & 3275.5~GeV& $m_{\tilde{\chi}_4^0}$   & 244.6~GeV& $A_\lambda$                  & 2078.0~GeV& $m_{\tilde{\chi}_4^0}$   & 313.2~GeV\\
$A_\kappa$            & -288.8~GeV& $m_{\tilde{\chi}_5^0}$   & 2020.9~GeV& $A_\kappa$                  & 521.3~GeV& $m_{\tilde{\chi}_5^0}$   & 2021.0~GeV\\
$A_t$                 & -3846.0~GeV& $m_{\tilde{\chi}_1^\pm}$   & 235.0~GeV& $A_t$                  & -3546.2~GeV& $m_{\tilde{\chi}_1^\pm}$   & 308.4~GeV\\
$M_1$                 & 150.0~GeV& $m_{\tilde{\chi}_2^\pm}$   & 2021.2~GeV& $M_1$                & 150.0~GeV& $m_{\tilde{\chi}_2^\pm}$   & 2021.2~GeV\\
$m_{B}$ & 149.9~GeV& $\Omega h^2$                & 0.122& $m_{B}$ & 101.5~GeV& $\Omega h^2$                & 0.130\\
$m_{C}$ & 16.7~GeV& $\sigma_{\rm eff}^{\rm SI}$                & $2.00\times 10^{-49}~{\rm cm}^2$& $m_{C}$ & 19.9~GeV& $\sigma_{\rm eff}^{\rm SI}$                & $1.22\times 10^{-48}~{\rm cm}^2$\\
$m_{\tilde S}$ & -52.5~GeV& $\sigma_n^{\rm SD}$                & $4.00\times 10^{-47}~{\rm cm}^2$& $m_{\tilde S}$ & -73.4~GeV& $\sigma_n^{\rm SD}$                & $5.13\times 10^{-50}~{\rm cm}^2$\\ \hline
\multicolumn{2}{c}{$V_{h_s}^S,~V_{h_s}^{\rm SM},~V_{h}^S,~V_{h}^{\rm SM}$} & \multicolumn{2}{c|}{~-0.999,~-0.004,~~0.004,~-0.999} & \multicolumn{2}{c}{$V_{h_s}^S,~V_{h_s}^{\rm SM},~V_{h}^S,~V_{h}^{\rm SM}$} & \multicolumn{2}{c}{~~0.999,~-0.025,~~0.025,~~0.999}\\
\multicolumn{2}{c}{$N_{11},~N_{12},~N_{13},~N_{14},~N_{15}$} & \multicolumn{2}{c|}{-0.001,~~0.000,~~0.001,~~0.006,~-0.999} & \multicolumn{2}{c}{$N_{11},~N_{12},~N_{13},~N_{14},~N_{15}$} & \multicolumn{2}{c}{~0.000,~~0.000,~~0.000,~~0.001,~-0.999}\\
\multicolumn{2}{c}{$N_{21},~N_{22},~N_{23},~N_{24},~N_{25}$} & \multicolumn{2}{c|}{~0.939,~-0.008,~~0.290,~-0.183,~-0.002} & \multicolumn{2}{c}{$N_{21},~N_{22},~N_{23},~N_{24},~N_{25}$} & \multicolumn{2}{c}{~0.978,~-0.004,~~0.186,~-0.094,~~0.000}\\
\multicolumn{2}{c}{$N_{31},~N_{32},~N_{33},~N_{34},~N_{35}$} & \multicolumn{2}{c|}{-0.077,~~0.024,~~0.699,~~0.711,~~0.005} & \multicolumn{2}{c}{$N_{31},~N_{32},~N_{33},~N_{34},~N_{35}$} & \multicolumn{2}{c}{-0.065,~~0.023,~~0.701,~~0.710,~~0.001}\\
\multicolumn{2}{c}{$N_{41},~N_{42},~N_{43},~N_{44},~N_{45}$} & \multicolumn{2}{c|}{-0.334,~-0.031,~~0.654,~-0.678,~-0.003} & \multicolumn{2}{c}{$N_{41},~N_{42},~N_{43},~N_{44},~N_{45}$} & \multicolumn{2}{c}{-0.198,~-0.033,~~0.688,~-0.697,~~0.000}\\
\multicolumn{2}{c}{$N_{51},~N_{52},~N_{53},~N_{54},~N_{55}$} & \multicolumn{2}{c|}{~0.001,~-0.999,~-0.006,~~0.040,~~0.000} & \multicolumn{2}{c}{$N_{51},~N_{52},~N_{53},~N_{54},~N_{55}$} & \multicolumn{2}{c}{~0.001,~-0.999,~-0.008,~~0.040,~~0.000}\\ \hline
\multicolumn{2}{c}{ Annihilations }                                                                                  & \multicolumn{2}{c|}{Fractions [\%]} & \multicolumn{2}{c}{Annihilations}                                                                                  & \multicolumn{2}{c}{Fractions [\%]}                                                                                  \\
\multicolumn{2}{c}{$\tilde{\chi}_1^0\tilde{\chi}_1^0 \to A_s A_s  $} & \multicolumn{2}{l|}{99.8}        & \multicolumn{2}{c}{$\tilde{\chi}_1^0\tilde{\chi}_1^0 \to h_s A_s / A_s A_s $} & \multicolumn{2}{l}{70.0~/~30.0}        \\ \hline
\multicolumn{2}{c}{ Decays }                                                                                  & \multicolumn{2}{c|}{Branching ratios [\%]} & \multicolumn{2}{c}{Decays}                                                                                  & \multicolumn{2}{c}{Branching ratios [\%]}                                                                                  \\
\multicolumn{2}{l}{$h_s \to A_sA_s/\tilde{\chi}_1^0\tilde{\chi}_1^0$} & \multicolumn{2}{l|}{77.8~/~22.2}        & \multicolumn{2}{l}{$h_s \to A_sA_s$} & \multicolumn{2}{l}{100.0}        \\
\multicolumn{2}{l}{$h \to A_sA_s/\tilde{\chi}_1^0\tilde{\chi}_1^0/\cdots$} & \multicolumn{2}{l|}{10.0~/~1.0~/~$\cdots$}        & \multicolumn{2}{l}{$h \to A_sA_s/\cdots$} & \multicolumn{2}{l}{9.6~/~$\cdots$}        \\
\multicolumn{2}{l}{$H \to b\bar b/\tau\bar\tau/\cdots$} & \multicolumn{2}{l|}{70.1~/~13.1~/~$\cdots$}        & \multicolumn{2}{l}{$H \to b\bar b/\tau\bar\tau/\cdots$} & \multicolumn{2}{l}{71.8~/~13.4~/~$\cdots$}        \\
\multicolumn{2}{l}{$A_s \to b\bar b/\tau\bar\tau  $} & \multicolumn{2}{l|}{88.8~/~9.9}        & \multicolumn{2}{l}{$A_s \to b\bar b/\tau\bar\tau  $} & \multicolumn{2}{l}{88.4~/~10.2}        \\
\multicolumn{2}{l}{$\tilde{\chi}_2^0 \to \tilde{\chi}_1^0 A_s$} & \multicolumn{2}{l|}{100.0}        & \multicolumn{2}{l}{$\tilde{\chi}_2^0 \to \tilde{\chi}_1^0 A_s$} & \multicolumn{2}{l}{100.0}        \\
\multicolumn{2}{l}{$\tilde{\chi}_3^0 \to \tilde{\chi}_2^0 Z$} & \multicolumn{2}{l|}{98.1}        & \multicolumn{2}{l}{$\tilde{\chi}_3^0 \to  \tilde{\chi}_2^0 h / \tilde{\chi}_2^0 Z $} & \multicolumn{2}{l}{4.8~/~95.2}        \\
\multicolumn{2}{l}{$\tilde{\chi}_4^0 \to \tilde{\chi}_2^0  Z$} & \multicolumn{2}{l|}{98.5}        & \multicolumn{2}{l}{$\tilde{\chi}_4^0 \to \tilde{\chi}_2^0 h / \tilde{\chi}_2^0 Z$} & \multicolumn{2}{l}{81.2~/~18.8}        \\
\multicolumn{2}{l}{$\tilde{\chi}_5^0 \to \tilde{\chi}_3^0 h / \tilde{\chi}_3^0 Z / \tilde{\chi}_4^0 h/ \tilde{\chi}_4^0 Z/ \tilde{\chi}_{1}^\pm W^\mp $} & \multicolumn{2}{l|}{8.2~/~15.5~/~14.7~/~7.6~/~47.6}        & \multicolumn{2}{l}{$\tilde{\chi}_5^0 \to  \tilde{\chi}_3^0 h / \tilde{\chi}_3^0 Z / \tilde{\chi}_4^0 h/ \tilde{\chi}_4^0 Z/ \tilde{\chi}_{1}^\pm W^\mp $} & \multicolumn{2}{l}{8.4~/~15.5~/~13.7~/~7.3~/~47.7}        \\
\multicolumn{2}{l}{$\tilde{\chi}_1^\pm \to \tilde{\chi}_2^0 W^\pm  $} & \multicolumn{2}{l|}{99.5}        & \multicolumn{2}{l}{$\tilde{\chi}_1^\pm \to \tilde{\chi}_1^0 W^\pm $} & \multicolumn{2}{l}{100.0}        \\
\multicolumn{2}{l}{$\tilde{\chi}_2^\pm \to \tilde{\chi}_{3}^0 W^\pm /\tilde{\chi}_{4}^0 W^\pm  /\tilde{\chi}_{1}^\pm h /\tilde{\chi}_1^\pm Z$} & \multicolumn{2}{l|}{23.8~/~22.2~/~23.8~/~23.7}        & \multicolumn{2}{l}{$\tilde{\chi}_2^\pm \to \tilde{\chi}_{3}^0 W^\pm /\tilde{\chi}_{4}^0 W^\pm  /\tilde{\chi}_{1}^\pm h /\tilde{\chi}_1^\pm Z$} & \multicolumn{2}{l}{23.8~/~23.3~/~23.9~/~23.6}        \\ \hline
\multicolumn{2}{c}{$R$ value} & \multicolumn{2}{c|}{0.398}        & \multicolumn{2}{c}{$R$ value} & \multicolumn{2}{c}{0.295}        \\ \hline \hline
\end{tabular}}
\end{table}

The characteristic features of surviving samples are exemplified by two benchmark points detailed in Table~\ref{tab:BP-150}. 
Both have low $M_1$ and $\mu_{\rm tot}$, and consequently the DM acquires small but non-negligible Bino and Higgsino admixtures, especially in the $h_1$ case. These admixtures strengthen the destructive interference among the contributions to $\sigma^{\rm SI}_{\tilde\chi^0_1,N}$, thereby suppressing the effective cross-section $\sigma_{\rm eff}^{\rm SI}$ below the LZ 2024 limit. In addition, the sparticle spectra yield multi-step decay chains with several competing modes (often involving a light $A_s$), which weakens current LHC sensitivities and allows these benchmarks to evade existing searches.

\subsection{Subtleties of LHC restrictions}\label{sec:Subtleties}

The results summarized in Table~\ref{tab:h1h2-150} might suggest that the LHC constraints are largely inconsequential for light DM scenarios. This impression is misleading: depending on the sparticle mass spectrum, LHC searches can impose very strong limits on such scenarios. 
To illustrate this point, we now consider an alternative configuration in which the Bino is sufficiently heavy such that the sparticle spectrum exhibits the hierarchy  $m_{\tilde{S}} \ll m_{\tilde{H}} \ll M_1$ ($\tilde{S} < \tilde{H} < \tilde{B}$). Compared to the previously studied configuration, this hierarchy substantially simplifies the Higgsino decay patterns, which typically strengthens the sensitivity of LHC constraints. Specifically, we repeated the analysis strategy outlined previously, fixing 
$M_1 = 2~{\rm TeV}$. Comprehensive exploratory scans of the GNMSSM parameter space and careful examination of the acquired samples indicate that all samples with Higgsino masses below approximately 900 GeV are excluded by LHC experiments. This issue is discussed extensively below.

When characterizing light DM scenarios in the heavy Bino case, we followed the same procedure as described in Section~\ref{sec:RS}, with the sole modification being that the Higgsino mass was scanned in the range $850~{\rm GeV}$ to $1200~{\rm GeV}$. 
This scan initially produced 24842 samples for the $h_1$ scenario and 24902 samples for the $h_2$ scenario. After applying LZ 2024 constraints, these numbers were substantially reduced to 2885 and 139, respectively. The LHC simulation further narrowed the surviving sets to 1217 for the $h_1$ scenario and only 15 for the $h_2$ scenario by the nominal requirement $R<1$.
For completeness, we also adopt a loose LHC criterion $R\le 1.5$ to account for implementation uncertainties in the recasting (see discussion below). Following this analysis strategy, we documented the updated sampling results in Table~\ref{tab:h1h2-2000}, where the values inside the parentheses apply to the loose case.

\begin{table}[t]
    \caption{A series of results similar to those of Table \ref{tab:h1h2-150}, but differing in two regards. 1) $M_1=2~\rm{TeV}$. 2) In the second row of each scenario, the first numbers correspond to those further satisfying the LHC constraints (i.e., the typical $R<1$ results), while the values inside the parentheses apply to the loose case ($R\le 1.5$). \label{tab:h1h2-2000}}
    \centering
    \vspace{0.3cm}
    \resizebox{0.85\textwidth}{!}{
\begin{tabular}{c|c|c|c|c|c}
\hline\hline
\multicolumn{1}{c|}{\textbf{Scenario}} & \multicolumn{1}{c|}{\textbf{Sample Size}} & {$\tilde{\chi}_1^0 \tilde{\chi}_1^0  \to  h_s  h_s$} & \multicolumn{1}{c|}{$\tilde{\chi}_1^0 \tilde{\chi}_1^0  \to  A_s  A_s$} & \multicolumn{1}{c|}{$\tilde{\chi}_1^0 \tilde{\chi}_1^0  \to  h_s  A_s$} & \multicolumn{1}{c}{$\tilde{\chi}_1^0 \tilde{\chi}_1^0  \to  f  \bar f$} \\ \hline
\multirow{2}{*}{$h_1$} & 2885 & $0$ & $97.6\%$ & $1.7\%$ & $0.6\%$ \\ \cline{2-6}
 & 1217~(2117) & $0~(0)$ & $98.4\%~(97.8\%)$ & $1.2\%~(1.7\%)$ & $0.4\%~(0.5\%)$ \\ \hline
\multirow{2}{*}{$h_2$} & 139 & $3.6\%$ & $8.6\%$ & $86.3\%$ & $1.4\%$ \\ \cline{2-6}
 & 15~(123) & $0~(4.1\%)$ & $0~(8.9\%)$ & $100.0\%~(85.4\%)$ & $0~(1.6\%)$ \\ \hline
\end{tabular}}
\end{table}

Compared with the  $M_1=150~\rm{GeV}$ case (Table~\ref{tab:h1h2-150}), significantly fewer samples satisfy the LZ 2024 constraints when $M_1=2~\rm{TeV}$. The primary reason is that for large $M_1$ and $\mu_{\rm tot}$, the DM becomes an almost pure Singlino with negligible Bino and Higgsino admixtures. As discussed earlier, small Bino and  $\tilde{H}_d$ components are crucial for facilitating destructive interference among different contributions to $\sigma^{\rm SI}_{\tilde\chi^0_1,N}$.  
Specifically, their markedly stronger coupling to the SM-like Higgs boson, relative to that of the Singlino, can generate sizable scattering amplitudes that interfere destructively with the pure-Singlino amplitude. When these admixtures are driven to (nearly) zero by $M_1=2~\rm{TeV}$ and $\mu_{\rm tot} \gtrsim 900~{\rm GeV}$, this interference leverage is lost and the cancellation is greatly weakened, thereby impeding effective suppression of the scattering cross-section. Consequently, most samples fail to satisfy the LZ 2024 constraints and are thus excluded even when $\lambda$ is as low as 0.01. Given that this effect is particularly pronounced in the $h_2$ scenario, resulting in its severely limited viable sample size, we henceforth focus on the $h_1$ scenario.

In the heavy-Bino case, the dominant annihilation channel remains $\tilde{\chi}_1^0 \tilde{\chi}_1^0 \to A_s A_s$, with parameter-space distributions qualitatively similar to those at  $M_1=150~{\rm GeV}$ but with much lower sample density. Accordingly, the physical interpretations established in the  $M_1=150~{\rm GeV}$ analysis remain fully applicable, and we refrain from reiterating the detailed arguments here. One salient feature, however, warrants emphasis: the weakened cancellation among different contributions to $\sigma^{\rm SI}_{\tilde\chi^0_1,N}$ causes $\sigma_{\rm eff}^{\rm SI}$ to be substantially elevated. As a direct consequence, surviving samples tend to cluster near the LZ 2024 exclusion boundary, and configurations with very small $\sigma_{\rm eff}^{\rm SI}$ are considerably rarer compared to the $M_1=150~{\rm GeV}$ case shown in Fig.~\ref{fig:150-sigmaSI-DM-lambda-kappa}.

\begin{figure}[t]
\centering 
{\tiny
\includegraphics[scale=0.5]{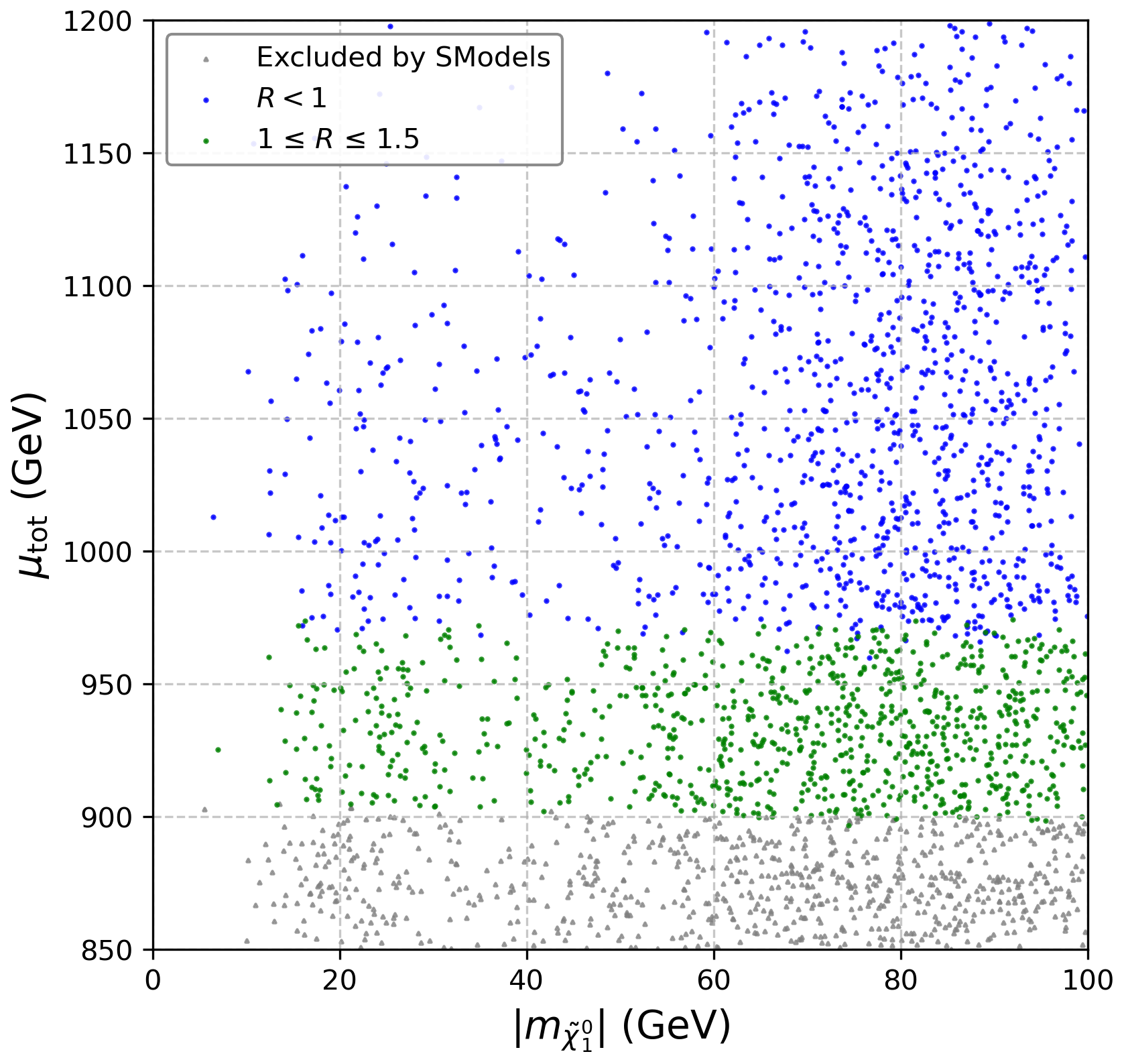} \quad\quad\quad
\includegraphics[scale=0.5]{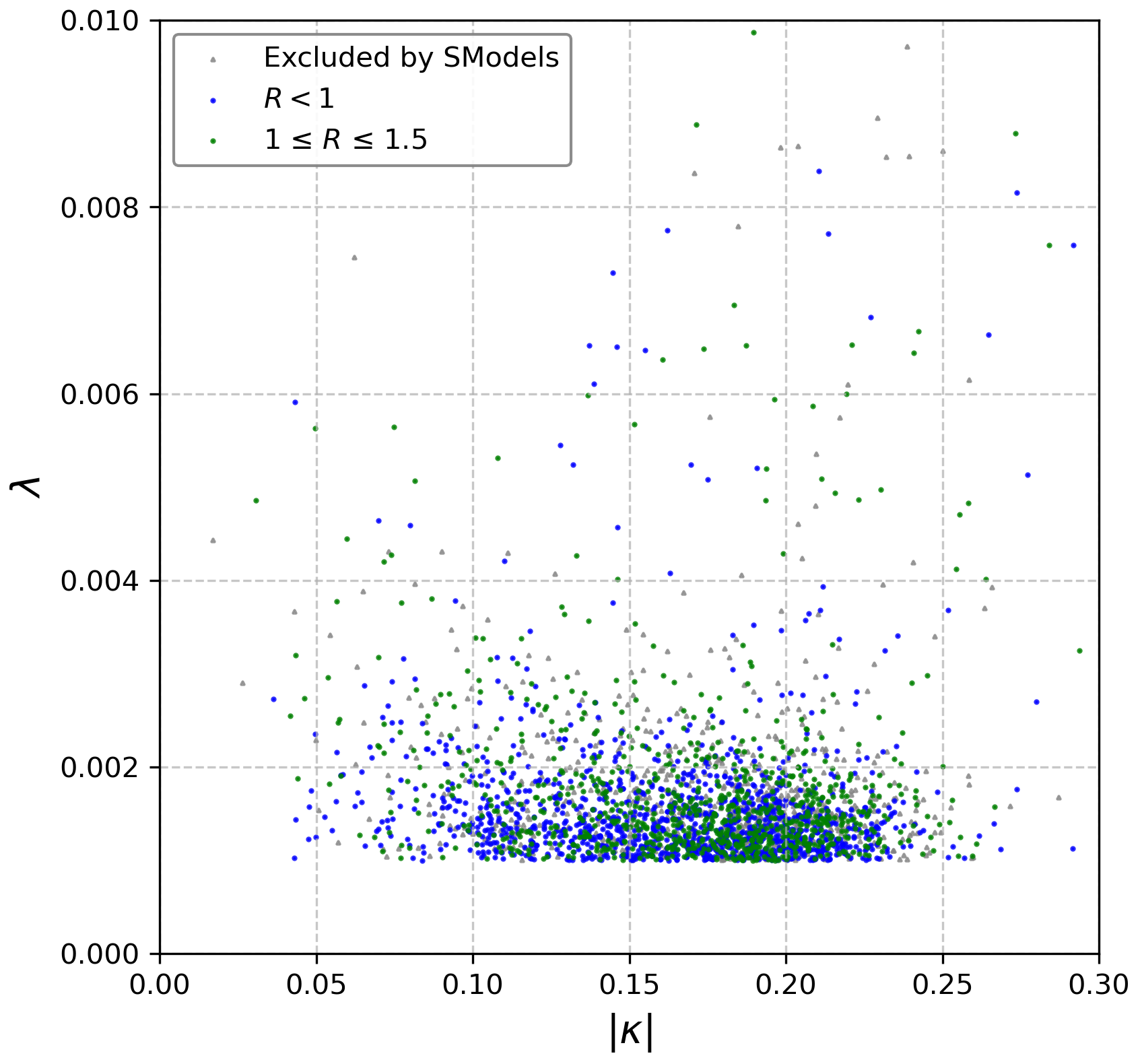} \\[-0.4cm]
}
\caption{Similar to Fig.~\ref{fig:LHC-h1h2-150}, but for $M_1=2~\rm{TeV}$ in the $h_1$ scenario. Samples with  $R \geq 1$ are further categorized: green markers ($1 \leq R \leq 1.5$) represent those satisfying the loose LHC constraint criterion, suggesting potential compatibility with LHC data once theoretical and experimental uncertainties are properly accounted for. These uncertainties arise from multiple sources, including sparticle production cross-section calculations and involved Monte Carlo simulations. \label{fig:LHC-h1-2000}}
\end{figure}
 
On the collider side, the adjustable sparticle spectrum consists of $\tilde{\chi}^0_{2,3}$ and charginos $\tilde{\chi}^\pm_1$, with their masses approximated by $|m_{\tilde{\chi}^0_{2,3}}| \simeq m_{\tilde{\chi}^\pm_1} \simeq \mu_{\rm tot}$, when $M_1=2~{\rm TeV}$.
Within the simplified mass hierarchy, the dominant decay channels for these particles are $\tilde{\chi}^0_{2,3}, \tilde{\chi}^\pm_1 \rightarrow \tilde{\chi}^0_1 + W/Z/h$, as exemplified by benchmark point {\bf BP3} in Table \ref{tab:BP-2000}. 
These signatures closely resemble those targeted by the ATLAS search for charginos and neutralinos decaying to fully hadronic final states in the MSSM, as documented in Ref.~\cite{ATLAS:2021yqv} (\texttt{ATLAS-2108-07586}). 
Critically, in light DM scenarios, this analysis sets a lower mass limit of approximately $\sim 900~{\rm GeV}$ on the NLSP mass, as illustrated in the left panel of Fig.~\ref{fig:LHC-h1-2000}. The right panel further shows that surviving points concentrate in narrow bands, $\lambda \in (0.001, 0.01)$ and $|\kappa| \in (0.05, 0.25)$.  

An important technical note concerns the $R \le 1.5$ exclusion criterion, which was adopted to account for implementation uncertainties encountered during the recasting of the \texttt{ATLAS-2108-07586} analysis into the \texttt{CheckMATE-2.0.37} framework~\cite{Kim:2015wza, Dercks:2016npn}. Consequently, interpreting samples yielding  $R \le 1.5$ as phenomenologically consistent with current LHC experiments is justified in this context. The resulting samples with  $R \le 1.5$ are depicted by the green-shaded region in Fig.~\ref{fig:LHC-h1-2000}. 

\begin{table}[t]
\caption{A list of results similar to Table \ref{tab:BP-150}, but with $M_1=2~{\rm TeV}$. \label{tab:BP-2000}}\
\centering
\resizebox{1\textwidth}{!}
{
\begin{tabular}{crcr|crcr}
\hline \hline
\multicolumn{4}{c|}{\bf Benchmark Point BP3~~~($h_1$ scenario)}                                                                                                & \multicolumn{4}{c}{\bf Benchmark Point BP4~~~($h_2$ scenario)}                                                                                                \\ \hline
$\lambda$             & 0.001     & $m_{h_s}$                & 192.1~GeV & $\lambda$             & 0.001& $m_{h_s}$                & 94.4~GeV\\
$\kappa$              & 0.186     & $m_{h}$                & 125.3~GeV & $\kappa$              & -0.076& $m_{h}$               & 125.2~GeV\\
$\delta$    & 0.923          & $m_{H}$ & 2550.0~GeV & $\delta$    & 0.721& $m_{H}$ & 1677.7~GeV\\
$\tan{\beta}$         & 29.51     & $m_{A_s}$                  & 10.0~GeV & $\tan{\beta}$         & 17.89& $m_{A_s}$                  & 15.7~GeV\\
$\mu$                 & 968.1~GeV & $m_{\tilde{\chi}_1^0}$                  & 35.0~GeV & $\mu$                 & 967.6~GeV& $m_{\tilde{\chi}_1^0}$                  & -85.4~GeV\\
$\mu_{\rm eff}$                 & 0.349~GeV & $m_{\tilde{\chi}_2^0}$                  & 981.8~GeV & $\mu_{\rm eff}$                 & 0.154~GeV& $m_{\tilde{\chi}_2^0}$                  & 979.1~GeV\\
$\mu_{\rm tot}$ & 968.5~GeV & $m_{\tilde{\chi}_3^0}$                & -986.7~GeV & $\mu_{\rm tot}$ & 967.8~GeV& $m_{\tilde{\chi}_3^0}$                & -985.2~GeV\\
$A_\lambda$            & 2182.2~GeV  & $m_{\tilde{\chi}_4^0}$   & 2006.6~GeV & $A_\lambda$                  & 4923.6~GeV& $m_{\tilde{\chi}_4^0}$   & 2006.8~GeV\\
$A_\kappa$            & 645.3~GeV  & $m_{\tilde{\chi}_5^0}$   & 2021.2~GeV & $A_\kappa$                  & -640.3~GeV& $m_{\tilde{\chi}_5^0}$   & 2021.3~GeV\\
$A_t$                 & 2694.7~GeV & $m_{\tilde{\chi}_1^\pm}$   & 984.0~GeV & $A_t$                  & -3493.0~GeV& $m_{\tilde{\chi}_1^\pm}$   & 981.8~GeV\\
$M_1$                 & 2000.0~GeV          & $m_{\tilde{\chi}_2^\pm}$   & 2021.0~GeV & $M_1$                & 2000.0~GeV& $m_{\tilde{\chi}_2^\pm}$   & 2021.2~GeV\\
$m_{B}$ & 194.6~GeV & $\Omega h^2$                & 0.136 & $m_{B}$ & 96.4~GeV& $\Omega h^2$                & 0.127\\
$m_{C}$ & 25.5~GeV & $\sigma_{\rm eff}^{\rm SI}$                & $1.90\times 10^{-48}~{\rm cm}^2$ & $m_{C}$ & 22.7~GeV& $\sigma_{\rm eff}^{\rm SI}$                & $2.61\times 10^{-48}~{\rm cm}^2$\\
$m_{\tilde S}$ & 35.1~GeV & $\sigma_n^{\rm SD}$                & $3.72\times 10^{-53}~{\rm cm}^2$ & $m_{\tilde S}$ & -85.4~GeV& $\sigma_n^{\rm SD}$                & $3.93\times 10^{-53}~{\rm cm}^2$\\ \hline
\multicolumn{2}{c}{$V_{h_s}^S,~V_{h_s}^{\rm SM},~V_{h}^S,~V_{h}^{\rm SM}$} & \multicolumn{2}{c|}{~-0.999,~-0.015,~~0.015,~-0.999} & \multicolumn{2}{c}{$V_{h_s}^S,~V_{h_s}^{\rm SM},~V_{h}^S,~V_{h}^{\rm SM}$} & \multicolumn{2}{c}{~~0.999,~-0.036,~~0.036,~~0.998}\\
\multicolumn{2}{c}{$N_{11},~N_{12},~N_{13},~N_{14},~N_{15}$} & \multicolumn{2}{c|}{~0.000,~~0.000,~~0.000,~~0.000,~~1.000} & \multicolumn{2}{c}{$N_{11},~N_{12},~N_{13},~N_{14},~N_{15}$} & \multicolumn{2}{c}{~0.000,~~0.000,~~0.000,~~0.000,~~0.999}\\
\multicolumn{2}{c}{$N_{21},~N_{22},~N_{23},~N_{24},~N_{25}$} & \multicolumn{2}{c|}{-0.031,~~0.054,~-0.707,~~0.704,~~0.000} & \multicolumn{2}{c}{$N_{21},~N_{22},~N_{23},~N_{24},~N_{25}$} & \multicolumn{2}{c}{-0.032,~-0.056,~-0.707,~~0.704,~~0.000}\\
\multicolumn{2}{c}{$N_{31},~N_{32},~N_{33},~N_{34},~N_{35}$} & \multicolumn{2}{c|}{~0.010,~-0.018,~-0.707,~-0.707,~~0.000} & \multicolumn{2}{c}{$N_{31},~N_{32},~N_{33},~N_{34},~N_{35}$} & \multicolumn{2}{c}{~0.010,~-0.017,~-0.706,~-0.708,~~0.000}\\
\multicolumn{2}{c}{$N_{41},~N_{42},~N_{43},~N_{44},~N_{45}$} & \multicolumn{2}{c|}{-0.474,~-0.880,~-0.016,~~0.031,~~0.000} & \multicolumn{2}{c}{$N_{41},~N_{42},~N_{43},~N_{44},~N_{45}$} & \multicolumn{2}{c}{-0.476,~-0.879,~-0.016,~~0.031,~~0.000}\\
\multicolumn{2}{c}{$N_{51},~N_{52},~N_{53},~N_{54},~N_{55}$} & \multicolumn{2}{c|}{~0.880,~-0.472,~-0.025,~~0.050,~~0.000} & \multicolumn{2}{c}{$N_{51},~N_{52},~N_{53},~N_{54},~N_{55}$} & \multicolumn{2}{c}{~0.879,~-0.474,~-0.027,~~0.051,~~0.000}\\ \hline
\multicolumn{2}{c}{ Annihilations }                                                                                  & \multicolumn{2}{c|}{Fractions [\%]} & \multicolumn{2}{c}{Annihilations}                                                                                  & \multicolumn{2}{c}{Fractions [\%]}                                                                                  \\
\multicolumn{2}{c}{$\tilde{\chi}_1^0\tilde{\chi}_1^0 \to A_s A_s  $} & \multicolumn{2}{l|}{100.0}        & \multicolumn{2}{c}{$\tilde{\chi}_1^0\tilde{\chi}_1^0 \to h_s A_s / A_s A_s $} & \multicolumn{2}{l}{98.7~/~1.2}        \\ \hline
\multicolumn{2}{c}{ Decays }                                                                                  & \multicolumn{2}{c|}{Branching ratios [\%]} & \multicolumn{2}{c}{Decays}                                                                                  & \multicolumn{2}{c}{Branching ratios [\%]}                                                                                  \\
\multicolumn{2}{l}{$h_s \to A_sA_s/\tilde{\chi}_1^0\tilde{\chi}_1^0$} & \multicolumn{2}{l|}{86.1~/~13.9}        & \multicolumn{2}{l}{$h_s \to A_sA_s$} & \multicolumn{2}{l}{100.0}        \\
\multicolumn{2}{l}{$h \to A_sA_s/\cdots$} & \multicolumn{2}{l|}{10.4~/~$\cdots$}        & \multicolumn{2}{l}{$h \to A_sA_s/\cdots$} & \multicolumn{2}{l}{8.4~/~$\cdots$}        \\
\multicolumn{2}{l}{$H \to b\bar b/\tau\bar\tau$} & \multicolumn{2}{l|}{79.6~/~19.5}        & \multicolumn{2}{l}{$H \to b\bar b/\tau\bar\tau$} & \multicolumn{2}{l}{81.1~/~15.7}        \\
\multicolumn{2}{l}{$A_s \to b\bar b/\tau\bar\tau  $} & \multicolumn{2}{l|}{87.9~/~10.6}        & \multicolumn{2}{l}{$A_s \to b\bar b/\tau\bar\tau  $} & \multicolumn{2}{l}{89.5~/~9.6}        \\
\multicolumn{2}{l}{$\tilde{\chi}_2^0 \to \tilde{\chi}_1^0 h/\tilde{\chi}_1^0 Z$} & \multicolumn{2}{l|}{51.7~/~47.8}        & \multicolumn{2}{l}{$\tilde{\chi}_2^0 \to \tilde{\chi}_1^0 h / \tilde{\chi}_1^0 Z $} & \multicolumn{2}{l}{37.8~/~62.0}        \\
\multicolumn{2}{l}{$\tilde{\chi}_3^0 \to \tilde{\chi}_1^0 h/\tilde{\chi}_1^0 Z$} & \multicolumn{2}{l|}{49.3~/~50.4}        & \multicolumn{2}{l}{$\tilde{\chi}_3^0 \to \tilde{\chi}_1^0 h / \tilde{\chi}_1^0 Z $} & \multicolumn{2}{l}{63.5~/~36.5}        \\
\multicolumn{2}{l}{$\tilde{\chi}_4^0 \to \tilde{\chi}_2^0 h / \tilde{\chi}_3^0 Z / \tilde{\chi}_1^\pm W^\mp $}  & \multicolumn{2}{l|}{10.2~/~9.4~/~75.7}        & \multicolumn{2}{l}{$\tilde{\chi}_4^0 \to \tilde{\chi}_2^0 h / \tilde{\chi}_3^0 Z / \tilde{\chi}_1^\pm W^\mp$} & \multicolumn{2}{l}{9.9~/~9.4~/~76.1}        \\
\multicolumn{2}{l}{$\tilde{\chi}_5^0 \to \tilde{\chi}_2^0 h / \tilde{\chi}_2^0 Z / \tilde{\chi}_3^0 h / \tilde{\chi}_{3}^0 Z $} & \multicolumn{2}{l|}{44.2~/~4.6~/~4.6~/~42.4}        & \multicolumn{2}{l}{$\tilde{\chi}_5^0 \to \tilde{\chi}_2^0 h /  \tilde{\chi}_2^0 Z /  \tilde{\chi}_3^0 h /\tilde{\chi}_{3}^0 Z $} & \multicolumn{2}{l}{43.0~/~4.1~/~4.1~/~42.4}        \\
\multicolumn{2}{l}{$\tilde{\chi}_1^\pm \to \tilde{\chi}_1^0 W^\pm  $} & \multicolumn{2}{l|}{100.0}        & \multicolumn{2}{l}{$\tilde{\chi}_1^\pm \to \tilde{\chi}_1^0 W^\pm $} & \multicolumn{2}{l}{100.0}        \\
\multicolumn{2}{l}{$\tilde{\chi}_2^\pm \to \tilde{\chi}_{2}^0 W^\pm /\tilde{\chi}_{3}^0 W^\pm /\tilde{\chi}_1^\pm h/\tilde{\chi}_1^\pm Z$} & \multicolumn{2}{l|}{23.3~/~23.4~/~24.6~/~23.4}        & \multicolumn{2}{l}{$\tilde{\chi}_2^\pm \to \tilde{\chi}_{2}^0 W^\pm /\tilde{\chi}_{3}^0 W^\pm /\tilde{\chi}_1^\pm h/\tilde{\chi}_1^\pm Z$} & \multicolumn{2}{l}{23.4~/~23.7~/~24.1~/~23.7}        \\ \hline
\multicolumn{2}{c}{$R$ value} & \multicolumn{2}{c|}{0.963}        & \multicolumn{2}{c}{$R$ value} & \multicolumn{2}{c}{0.958}        \\ \hline \hline
\end{tabular}}
\end{table} 

Table~\ref{tab:BP-2000}  presents two benchmark points that provide further insights. Compared to Table~\ref{tab:BP-150}, the large $\mu_{\rm tot}$ value produces an almost pure Singlino DM, which precludes effective cancellation among different contributions to $\sigma^{\text{SI}}_{\tilde\chi^0_1,N}$, leading to surviving samples with moderately large $\sigma^{\text{SI}}_{\text{eff}}$. Moreover, the simpler decay topologies in the $\tilde S < \tilde H < \tilde B$  hierarchy  yield more detectable signatures at the LHC, resulting in notably more restrictive experimental limits.

\section{\label{conclusion}Conclusion}

Light supersymmetric DM with a mass of several tens of GeV warrants particular attention, as such light particles are challenging to realize within simple supersymmetric frameworks such as the MSSM or the $Z_3$-NMSSM. Through a comprehensive theoretical and numerical analysis incorporating relic density, LZ 2024, Higgs data, and LHC constraints, this study establishes the GNMSSM as a robust and phenomenologically viable framework capable of naturally accommodating light supersymmetric DM. The central virtue of the GNMSSM is the effective decoupling between interactions governing thermal relic abundance and those controlling direct detection, which is enabled by allowing $\mu$ and $\mu^\prime$ parameters to break the $Z_3$ symmetry of the $Z_3$-NMSSM and thereby disentangle $\lambda$ and $\kappa$. This feature broadens the viable parameter space and yields distinctive signatures that can help discriminate supersymmetric frameworks in future experiments. Our systemic parameter space exploration across two configurations —$M_1 = 150~\rm{GeV}$ and $M_1 = 2~\rm{TeV}$— yields the following key insights:
\begin{itemize}
\item \textbf{Diverse Self-Annihilation Mechanisms:} In the $h_1$ scenario, DM achieves the observed relic density primarily through the $\tilde{\chi}_1^0 \tilde{\chi}_1^0 \to A_s A_s$ channel, exhibiting pronounced mass-dependent behavior across three distinct regimes. In the low-mass region  ($\lesssim 30$ GeV), the $t$-channel dominates the annihilation process. The intermediate-mass region (30-60 GeV) features a delicate interplay between $t$- and $s$-channel contributions. In the high-mass region ($\gtrsim 60$ GeV), $s$-channel processes typically prevail through both resonant and non-resonant effects. In contrast, the $h_2$ scenario yields sufficient samples only for the  $M_1=150~{\rm GeV}$ configuration, where DM achieves the correct relic density primarily via  $\tilde{\chi}_1^0 \tilde{\chi}_1^0 \to h_s A_s$, with comparable $s$- and $t$-channel contributions in most parameter regions.
 
\item \textbf{Correlation Between Direct Detection and Naturalness:} For a heavy doublet-like Higgs $H$, the compatibility of $\sigma_{\rm eff}^{\rm SI}$ with LZ 2024 constraints critically depends on the strength of destructive interference among various contributions to $\sigma^{\rm SI}_{\tilde\chi^0_1,N}$. After incorporating constraints from LHC supersymmetry searches, this cancellation mechanism is substantially stronger at $M_1 = 150$ GeV than at $M_1 = 2$ TeV. Specifically, the former configuration can suppress  $\sigma_{\rm eff}^{\rm SI}$  down to $10^{-50}{\rm cm}^2$ while maintaining a larger viable parameter space, thereby more readily satisfying experimental constraints. Crucially, the $M_1 = 150$ GeV configuration permits $\mu_{\rm tot}$ values as low as $\sim 200~{\rm GeV}$, revealing an intrinsic consistency between direct detection constraints and theoretical naturalness.

\item \textbf{NLSP-Dependent LHC Constraints:} The identity of NLSP profoundly influences LHC constraints.
When  $M_1 = 150$ GeV, the NLSP is a Bino-dominated $\tilde{\chi}_2^0$. The resulting diverse decay modes and prolonged cascade decay chains degrade LHC search sensitivities, thereby allowing $\mu_{\rm tot}$ values as low as $\sim 200$ GeV without contradicting current experimental bounds. Consequently, the model preserves extensive parameter space for future LHC exploration: in the $h_1$ scenario,  $|m_{\tilde\chi^0_1}| \in (10, 100)~\mathrm{GeV}$, $\lambda \in (0., 0.05)$, and $|\kappa| \in (0.05, 0.3)$, while in the $h_2$ scenario, $|m_{\tilde\chi^0_1}| \in (50, 100)~\mathrm{GeV}$, $\lambda \in (0., 0.01)$, and $|\kappa| \in (0.03, 0.2)$. Both scenarios feature $m_{\tilde{\chi}_2^0} \simeq 150~{\rm GeV}$.  Conversely, when $M_1 = 2~\mathrm{TeV}$, the NLSP becomes Higgsino-dominated, leading to simplified decay topologies where Higgsino-like particles decay directly to the DM. LHC experiments then impose a stringent bound of $\mu_{\rm tot} \gtrsim 900~{\rm GeV}$ for such simplified signatures, substantially restricting the viable parameter space.

\end{itemize}

In summary, the GNMSSM circumvents the $Z_3$-NMSSM constraint $2|\kappa| < \lambda$ in predicting Singlino-dominated DM, thereby enabling effective decoupling between DM annihilation mechanisms and DM-nucleon scattering processes. Our analysis identifies two distinct mass hierarchies within the GNMSSM: $\tilde{S} < \tilde{B} < \tilde{H}$ and $\tilde{S} < \tilde{H} < \tilde{B}$, both capable of yielding phenomenologically viable DM candidates in the tens-of-GeV range. 
The $\tilde{S} < \tilde{B} < \tilde{H}$ hierarchy emerges as the preferred configuration, allowing $\mu_{\rm tot}$ values as low as $\sim 200~{\rm GeV}$ and naturally accommodating Singlino-dominated light DM with proper DM relic abundance. Moreover, the presence of accompanying light particles such as $A_s$ with characteristic LHC signatures provides additional handles for distinguishing between different supersymmetric models. By contrast, although the alternative $\tilde{S} < \tilde{H} < \tilde{B}$ hierarchy is theoretically simpler, it requires $\mu_{\rm tot} \gtrsim 900~{\rm GeV}$ based on LHC constraints, resulting in approximately $0.5\%$ fine-tuning in electroweak symmetry breaking.

By systematically exploring the previously uncharted $\tilde{S} < \tilde{B} < \tilde{H}$ parameter space, this work advances our understanding of supersymmetric DM in challenging mass regimes and provides concrete guidance for future experimental searches across multiple frontiers: direct detection experiments approaching $10^{-50}~{\rm cm^2}$ sensitivity, collider searches for electroweakinos with complex decay chains involving light singlet states, and indirect detection through secluded-sector annihilation products. We anticipate that forthcoming experiments will provide crucial tests of the mass hierarchy identified in this study, ultimately helping to establish the correct theoretical framework for describing DM.

\section*{Acknowledgements}
This work was supported by the National Natural Science Foundation of China (NNSFC) under grant No. 12575110. We thank LetPub (www.letpub.com.cn) for linguistic assistance and pre-submission expert review. The authors also gratefully acknowledge the valuable discussions and insights provided by the members of the China Collaboration of Precision Testing and New Physics(CPTNP).

\bibliographystyle{CitationStyle}
\bibliography{dm}

\end{document}